\documentclass[
aps, prb,
 floatfix,
 reprint,
 preprintnumbers,
 superscriptaddress,
 amsmath, amssymb,
 longbibliography,
]{revtex4-2}

\usepackage{float}

\newcommand*{\papertitle}{Thermopower in hBN/graphene/hBN superlattices}
\newcommand*{\affman}{Department of Physics and Astronomy, University of Manchester, M13 9PL, Manchester, UK}
\newcommand*{\affmandom}{Department of Materials, University of Manchester, M13 9PL, Manchester, UK}
\newcommand*{\affngi}{National Graphene Institute, University of Manchester, M13 9PL, Manchester, UK}
\newcommand*{\affecm}{Facultad de Ciencias Naturales y Matemáticas, Escuela Superior Politécnica del Litoral, ESPOL, Campus Gustavo Galindo, Km. 30.5 Vía Perimetral, P.O. Box 09-01-5863, 090902 Guayaquil, Ecuador}
\newcommand*{\affecd}{Center of Nanotechnology Research and Development (CIDNA), Escuela Superior Politécnica del Litoral, ESPOL, Campus Gustavo Galindo Km 30.5 Vía Perimetral, Guayaquil, Ecuador}
\newcommand*{\affbud}{Budapest University of Technology and Economics, Institute of Physics, Budapest, Hungary}
\newcommand*{\affkan}{Tim Taylor Department of Chemical Engineering, Kansas State University, Manhattan, Kansas 66506, USA}

\usepackage{graphicx}
\usepackage{bm}
\usepackage[dvipsnames]{xcolor}
\usepackage[breaklinks=true]{hyperref} 
\hypersetup{bookmarksnumbered, pdfpagemode=UseOutlines, pdfauthor={C.\ R.\ Anderson}, pdftitle={\papertitle}, 
pdfdisplaydoctitle, colorlinks=true, citecolor=blue, filecolor=blue, linkcolor=blue, urlcolor=blue}
\usepackage{verbatim}
\usepackage{textcomp}
\usepackage{units}

\begin{document}

\title{\papertitle}

\author{Victor H. \surname{Guarochico-Moreira}}
\thanks{These authors made an equal contribution and are equal first authors.}
\affiliation{\affman}
\affiliation{\affecm}
\affiliation{\affecd}

\author{Christopher R. \surname{Anderson}}
\thanks{These authors made an equal contribution and are equal first authors.}
\affiliation{\affman}

\author{Vladimir \surname{Fal'ko}}
\affiliation{\affman}
\affiliation{\affngi}

\author{Irina V. \surname{Grigorieva}}
\affiliation{\affman}
\affiliation{\affngi}

\author{Endre \surname{T\'{o}v\'{a}ri}}
\affiliation{\affngi}
\affiliation{\affbud}

\author{Matthew \surname{Hamer}}
\affiliation{\affman}
\affiliation{\affngi}

\author{Roman \surname{Gorbachev}}
\affiliation{\affman}
\affiliation{\affngi}

\author{Song \surname{Liu}}
\affiliation{\affkan}

\author{James H. \surname{Edgar}}
\affiliation{\affkan}

\author{Alessandro \surname{Principi}}
\affiliation{\affman}

\author{Andrey V. \surname{Kretinin}}
\thanks{Correspondance to andrey.kretinin@manchester.ac.uk and ivan.veramarun@manchester.ac.uk}
\affiliation{\affman}
\affiliation{\affngi}
\affiliation{\affmandom}

\author{Ivan J. \surname{Vera-Marun}}
\thanks{Correspondance to andrey.kretinin@manchester.ac.uk and ivan.veramarun@manchester.ac.uk}
\affiliation{\affman}
\affiliation{\affngi}

\date{\today}

\begin{abstract}
Thermoelectric effects are highly sensitive to the asymmetry in the density of states around the Fermi energy and can be exploited as probes of the electronic structure. We experimentally study thermopower in high-quality monolayer graphene, within heterostructures consisting of complete hBN encapsulation and 1D edge contacts, where the graphene and hBN lattices are aligned. 
When graphene is aligned to one of the hBN layers, we demonstrate the presence of additional sign reversals in the thermopower as a function of carrier density, directly evidencing the presence of the moir\'{e} superlattice. We show that the temperature dependence of the thermopower enables the assessment of the role of built-in strain variation and van Hove singularities and hints at the presence of Umklapp electron-electron scattering processes. As the thermopower peaks around the neutrality point, this allows to probe the energy spectrum degeneracy. 
Further, when graphene is double-aligned with the top and bottom hBN crystals, the thermopower exhibits features evidencing multiple cloned Dirac points caused by the differential super-moir\'{e} lattice. 
For both cases we evaluate how well the thermopower agrees with Mott’s equation. 
Finally, we show the same superlattice device can exhibit a temperature-driven thermopower reversal from positive to negative and vice versa, by controlling the carrier density. 
The study of thermopower provides an alternative approach to study the electronic structure of 2D superlattices, whilst offering opportunities to engineer the thermoelectric response on these heterostructures. 
\end{abstract}

\maketitle
\footnotetext[1]{These authors contributed equally and are equal first authors.}

\section*{Introduction}

The knowledge of the thermoelectric properties of a given material system is technologically significant for application in temperature sensing, waste heat harvesting and solid-state cooling \cite{He_Tritt_review}. From a fundamental perspective, it complements electronic transport studies of low-dimensional systems and provides additional information on the electronic spectrum \cite{Houtent1992}, scattering mechanisms \cite{Ghahari2016}, and possible many-body effects \cite{Mahapatra2020, Paul2022, Ghawri2022}. Graphene is one of the low-dimensional materials with thoroughly studied thermoelectric properties, mainly because of the relative simplicity of the system and well-developed experimental techniques \cite{Graphene_TE_review_Xu_Li_Duan}. Most importantly, graphene provides the opportunity to validate the so-called Mott's thermopower equation devised for a degenerate conductor in a diffusive regime \cite{cutler_observation_1969}. Early experiments with disordered monolayer graphene demonstrated an excellent agreement with the formula over a wide range of temperatures and charge carrier concentrations and allowed the study of the magneto-thermopower and the Nernst effect \cite{zuev_thermoelectric_2009, wei_anomalous_2009, checkelsky_thermopower_2009}. Later, these results were confirmed using higher quality material \cite{duan_high_2016, Ghahari2016}. Also, some peculiarities in the thermopower temperature dependence attributed to the parabolic spectrum of the bilayer graphene have been reported \cite{Nam_bilayer_TE, Wang_bilayer_TE}.

Further development of fabrication techniques improved the quality of the graphene devices \cite{Frisenda_fab_review} and provided the opportunity to study graphene systems beyond the classic diffusive regime. Thereby Mott's equation was found to breakdown in a monolayer in the hydrodynamic regime \cite{Ghahari2016} and in strongly-correlated states of a magic angle twisted bilayer \cite{Mahapatra2020, Paul2022, Ghawri2022}.

According to Mott's equation \cite{cutler_observation_1969}, the thermopower, $S$, contains information on the electronic spectrum and scattering processes hidden inside the energy-dependent conductivity $\sigma(E)$ measured at the Fermi level $E_{\text{F}}$,
\begin{equation} 
S(T, E_{\text{F}}) = -\frac{\pi^2k_B^2T}{3e} \frac{1}{\sigma(E)} \frac{\partial \sigma(E)}{\partial E}\Bigr|_{E_F},
\label{eq:DifferentialMott}
\end{equation}
where the conductivity $\sigma(E)=e^{2}\upsilon_{\text{F}}^{2}D(E)\tau(E)/2$, $\upsilon_{\text{F}}$ is the Fermi velocity, $D(E)$ is the density of states (DoS), $\tau(E)$ is the energy-dependent scattering time, $T$ is the device (electronic) temperature, $k_{B}$ is the Boltzmann constant and $e$ is the elementary charge. On the one hand, the influence of the DoS is not always obvious in thermoelectric experiments since the monolayer and bilayer graphene energy dispersion is featureless, and the strong interaction effects obscure the rich band structure of the twisted bilayer system. On the other, the energy-dependent scattering processes and variations in the effective mass can be dominant factors determining the value of the thermopower. Therefore, a careful analysis of the thermopower's temperature and energy dependencies has to be used to separate different contributory factors.

In the present work we studied the thermopower of monolayer graphene aligned with hexagonal boron nitride (hBN) \cite{Woods_commensurate_graphene_hBN}. In this system, the charge carriers remain uncorrelated yet the energy spectrum is strongly modified with the periodic moir\'{e} potential, which is manifested in multiple "cloned" Dirac points \cite{ponomarenko_cloning_2013, hunt_massive_2013, dean_hofstadter/s_2013}. In our experiments the temperature-dependent thermopower appears to be a rich source of information about charge homogeneity, built-in strain variation, energy degeneracy of the Dirac fermions near the neutrality point, and it hints at the presence of the moir\'{e}-induced Umklapp electron-electron scattering \cite{wallbank_excess_2018}. We also show that the Mott's equation (Eq.~\ref{eq:DifferentialMott}) captures well the features in the thermopower associated with the non-monotonic DoS even in the more complex case of a double-aligned hBN/graphene/hBN heterostructure with differential super-moir\'{e} Dirac points \cite{wang_composite_2019}; however, it fails to describe its value accurately.

\begin{figure*}[tbp]
	\includegraphics*[angle=0, trim=0mm 0mm 0mm 0mm, width=1\textwidth]{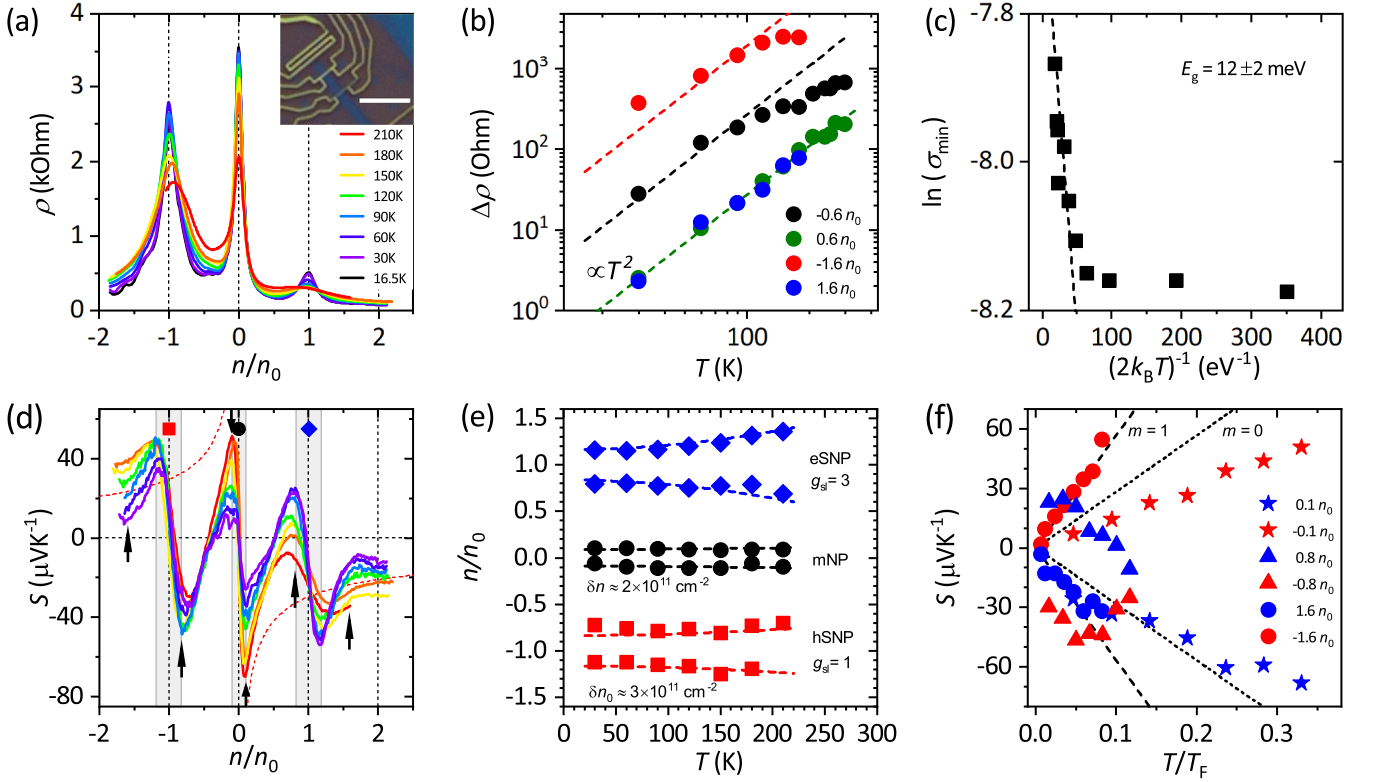}
	\caption{
		\label{fig:concept}
		\textbf{Charge transport and thermopower of the encapsulated single-aligned graphene/hBN device.} 
  (a) Resistivity as a function of normalised carrier density, ${n/n_0}$, at different temperatures $T$. The secondary neutrality points appear at carrier density $-n_0$ and $n_0$ ($n_{0}\approx 2.2\times10^{-12}$~cm\textsuperscript{-2}). Inset shows an optical microscopy image of the device. Scale bar is 5\,\textmu m. 
  (b) Temperature dependence of the excess resistivity $\Delta\rho$ relative to that at 16.5\,K at different fixed carrier densities. The dashed lines depicts the quadratic temperature dependence expected for the electron-electron Umklapp scattering from the edges of the moir\'{e} mini Brillouin zone.
  (c) Arrhenius plot of the conductivity at the main neutrality point $\sigma_{\text{min}}$  used to estimate the width of the apparent band gap ($E_{\text{g}}=12\pm2$~eV) extracted from the slope of the activation dependence (dashed line). 
  (d) Thermopower, ${S}$, as a function of the normalised carrier density, ${n/n_0}$, at the same temperatures as in (a). The dashed red curve is the prediction from Eq.~\ref{eq:TF_Mott} for unaligned graphene with an energy-independent scattering time ($m=0$) at $T=210$~K. The three vertical shaded bands mark the sign reversal regions associated with the corresponding charge neutrality points.
  (e) Position of the thermopower minima and maxima associated with each of the charge reversal regions shown in (d) and marked by the same data point symbols, plotted as a function of temperature. Dashed curves represent the expected value of $\pm\Delta/2$ given by Eq.~\ref{eq:Delta} with the appropriate superlattice degeneracy $g_{\text{sl}}$ carrier density uncertainties $\delta n$ and $\delta n_{0}$.
  (f) Thermopower at selected carrier densities marked by arrows in (d) as a function of dimensionless temperature $T/T_{\text{F}}$. The dashed (dotted) line is the prediction from Eq.~\ref{eq:TF_Mott} for $m=1$ ($m=0$).}
\end{figure*}

\section*{Results and Discussion}

The device architecture used in this study is a 1D edge contacted monolayer graphene channel, fully encapsulated between thin (20--40~nm) crystals of hBN, with the graphene crystallographically aligned to one or both of them. 
The thermopower measurements were performed utilising the well-established device configuration with on-chip microscale thermometers and a heater element \cite{small_mesoscopic_2003}. However, we modified the device geometry to improve the accuracy of the thermopower measurements in the four-terminal configuration \cite{Guarochico-Moreira2022}. Namely, the thermometers' sensing strip also serves as a 1D potential contact for the device, as shown in the inset of Fig.~\ref{fig:concept}a. That self-aligned configuration minimises systematic errors in the measurements of thermopower, caused by the spatial separation of the thermometer and potential probes typical for earlier designs \cite{ghahari_enhanced_2016,zuev_thermoelectric_2009,wei_anomalous_2009,checkelsky_thermopower_2009,duan_high_2016,Nam_bilayer_TE,Wang_bilayer_TE,small_mesoscopic_2003}.

Charge transport measurements for the single-aligned device (low-temperature field-effect mobility $\sim3\times 10\textsuperscript{4}$~cm\textsuperscript{2}V\textsuperscript{-1}s\textsuperscript{-1}, mean free path $\leq400$~nm) presented in Fig.~\ref{fig:concept}(a) reveal features typical of a graphene/hBN superlattice.  The two secondary neutrality points (SNPs) are situated symmetrically with respect of the main neutrality point (mNP), in the hole (hSNP) and electron (eSNP) regimes, and occur at the carrier concentration $n=n_{0}\approx\pm2.2\times10^{12}$~cm\textsuperscript{-2}. These SNPs are induced by Bragg scattering of charge carriers from the weak moir\'{e} superlattice potential when the lowest electron or hole minibands are fully occupied \cite{ponomarenko_cloning_2013, yankowitz_emergence_2012, hunt_massive_2013, dean_hofstadter/s_2013}. In the case of monolayer graphene, full filling occurs at a density of ${n_0 = 2g_{\text{s}}g_{\text{v}}/(\sqrt{3}\lambda^2)}$, where $g_{\text{s}}=2$ and $g_{\text{v}}=2$ are spin and valley degeneracy, respectively, and $\lambda$ is the superlattice period. The experimental value of $n_{0}$ provides the estimate for the superlattice period (moir\'{e} wavelength) to be $\lambda\approx14.5$~nm, which is larger than the maximum value of of 13.8~nm expected for an ideally aligned graphene-hBN system. That can be explained by an additional built-in strain imposed onto the graphene crystal, which effectively reduces the lattice mismatch to around 1.73\% and allows for a longer period superlattice (strain $\sim$0.07\%) \cite{wang_composite_2019}. 

The charge transport measurements show a metallic-like behaviour of electrical conductivity in the regions between the neutrality points, with a resistivity monotonically increasing with temperature. The secondary Dirac peaks seemed to be smeared by the elevating temperature, while the main Dirac peak remains less affected, see Fig.~\ref{fig:concept}(a). A closer look at the temperature-dependent resistivity showed that at the carrier densities ${0.25n_0 < |n| < 0.8n_0}$ and $|n|>1.2n_{0}$ the excess resistivity, defined as $\Delta \rho = \rho(T) - \rho(T=16.5\text{K})$, increases quadratically with the temperature, as illustrated in Fig.~\ref{fig:concept}(b).  This temperature dependence is a signature of the electron-electron Umklapp scattering from the edges of the moir\'{e}-induced mini Brillouin zone \cite{wallbank_excess_2018} and commonly observed in similar graphene/hBN superlattice systems \cite{proximity_screening_Geim}. The deviation from the ${T^2}$ dependence, evident from Fig.~\ref{fig:concept}(b), is likely due to the thermal excitation of carriers with the opposite sign of effective mass.

A number of earlier works \cite{hunt_massive_2013, song_electron_2013, san-jose_spontaneous_2014, san-jose_electronic_2014, bokdam_band_2014, jung_origin_2015, moon_electronic_2014} suggested that the periodic moir\'{e}-induced strain breaks the lattice's inversion symmetry and results in the gap opening at the main and secondary charge neutrality points. The Arrhenius plot of the electrical conductivity at the main Dirac peak shown in Fig.~\ref{fig:concept}(c) revealed some activation-like behavior in a limited range of temperatures, corresponding to the apparent band gap of $E_{\text{g}}\approx12\pm2$~meV. Similar temperature dependence and the value of the gap has been observed at hSNP. Both values are in general agreement with those previously reported in the literature. However, as seen from Fig.~~\ref{fig:concept}(c), the conductivity saturates at low temperatures and grows only by a limited amount at $T>100$~K. As it has been argued \cite{gorbachev_detecting_2014}, the charge inhomogeneity (electron-hole puddles) and midgap states can obscure the transport gap and make it undetectable in transport experiments. This is likely the case here, since the value of charge fluctuations $\delta n$ extracted from the width of the mNP at low temperature \cite{kretinin_electronic_2014} was between $8\times10^{10}$~cm\textsuperscript{-2} and $2\times10^{11}$~cm\textsuperscript{-2}, which corresponds to an uncertainty of the  potential $\delta\mu = \hbar \nu_{F} \sqrt{\pi \delta n} \geq 33$~meV  \cite{martin_observation_2008}. 

The thermopower of our aligned graphene/hBN system measured as a function of the carrier density for different temperatures is shown in Figure~\ref{fig:concept}(d). As seen from the figure, its value exhibits sign reversal regions (marked by vertical shaded bands) each associated with the corresponding charge neutrality point at the carrier densities of 0 and ${\pm n_0}$. The extent of the charge reversal region is defined as the difference in the position of the local maximum and minimum of the thermopower and is dictated by the magnitude of the charge density fluctuations and by the density of the thermally excited intrinsic carriers \cite{duan_high_2016}. Essentially, the charge reversal width is directly proportional to the width of the Dirac peak, reflecting the sample homogeneity and the temperature of the carriers. The sign reversal region for the mNP is relatively narrow ($\pm 2\times 10^{11}$~cm\textsuperscript{-2}) and almost independent of temperature, while for the hSNP and eSNP, it is twice as wide at 30\,K ($\pm 4\times 10^{11}$~cm\textsuperscript{-2}) and widens even further with increasing temperature. Generally, the width of the sign reversal, $\Delta$, is twice the sum of the r.m.s. value of the charge inhomogeneity $\delta n$ and the density of thermally excited carriers $n_{\text{i}}(T)$ \cite{duan_high_2016, fang_carrier_2007}:
\begin{equation}
\label{eq:Delta}
    \frac{\Delta}{2} = \delta n + n_{\text{i}}(T) = \delta n + \frac{g \pi}{24}\left(\frac{k_{\text{B}}T}{\hbar \upsilon_{\text{F}}}\right)^{2}
\end{equation}
where $\upsilon_{\text{F}} $ is the Fermi velocity and $g$ is the total system degeneracy. 
The evolution of the sign reversal regions with the temperature is illustrated in Fig.~\ref{fig:concept}(e). Here the position of the thermopower minima and maxima associated with each of the Dirac peaks is plotted as a function of $T$ with the dashed curves representing the expected value of $\pm\Delta/2$ given by Eq.~\ref{eq:Delta}. 

For the mNP, the spectrum is four-fold degenerate ($g=g_{s}g_{v}=4$) with $\upsilon_{\text{F}} = 1\times10^{6}$~m s\textsuperscript{-1}. The position predicted by Eq.~\ref{eq:Delta}, shown by the black dashed curve in Fig.~\ref{fig:concept}(e), agrees well with the experimental data (black circles). The value of charge inhomogeneity from the curve fitting is $\delta n \approx 2 \times 10^{11}$~cm\textsuperscript{-2}, which is close to that found from the width of the main Dirac peak at 16.5~K.  At the mNP the thermally excited intrinsic charge carriers contribute only a small fraction to the overall value of $\Delta$ ($n_{\text{i}}(T)<0.2\delta n$) and it is beyond the resolution of the measurement, thus the width of the sign reversal appears virtually independent of $T$.

The situation is somewhat different for the secondary neutrality points. Firstly, the initial width of the secondary sign reversal regions, $\Delta$, is approximately two times larger and cannot be explained by the presence of the charge inhomogeneity alone (here we assumed $\delta n$ to be independent on the total carrier density $n$). A possible source of extra uncertainty in the secondary Dirac peaks position is the spatial non-uniformity of the alignment angle or variation of the built-in strain that leads to the uncertainty $\delta n_{0}$ of the filling concentration $n_{0}$. Secondly, the density of the thermally activated carriers at the SNPs, is different due to the additional degeneracy of the secondary Dirac points, $g_{\text{sl}}$, and the reduced value of the Fermi velocity $\upsilon_{F}=0.5\times10^{6}$~m/s \cite{yankowitz_emergence_2012,Yu2014}. To account for these differences in Eq.~\ref{eq:Delta}, $\delta n$ should be replaced with $\sqrt{\delta n^{2} + \delta n_{0}^{2}}$ and the total system degeneracy is now $g = g_{\text{s}}g_{\text{v}}g_{\text{sl}}$ with  $g_{\text{sl}}=1$ for hSNP and $g_{\text{sl}}=3$ for eSNP. Due to these changes in degeneracy and Fermi velocity, the width of the secondary sign reversals (and the width of the secondary Dirac peaks) are more sensitive to the changing temperature. As can be seen in Fig.~\ref{fig:concept}(e), the experimental position of the thermopower maxima and minima (blue and red symbols) are approximated well by Eq.~\ref{eq:Delta} (blue and red dashed curves), with the temperature dependence being more pronounced for the eSNP because of the three times higher degeneracy. The uncertainty of the filling concentration used to fit the above experimental data was found to be $\delta n_{0}\approx3\times10^{11}$~cm\textsuperscript{-2}. This value allows the estimation of the superlattice period deviation $\delta\lambda = \left(\nicefrac{\sqrt{3}\lambda^{3}}{16}\right)\delta n_{0}\approx \pm1$~nm, which translates to the variation of the built-in strain of $\pm0.11\%$ (here we assumed that the graphene is ideally aligned).

The temperature dependence of the thermopower, shown in Fig.~\ref{fig:concept}(d), is more complex than usually observed in a monolayer graphene samples \cite{ghahari_enhanced_2016,duan_high_2016} primarily due to the presence of the secondary Dirac cones and the van Hove singularities associated with them.  At certain carrier densities the $S(T)$ dependence is more straightforward and some information about the underlying carrier scattering mechanisms can be extracted from it. To obtain this information, we plotted the thermopower at selected carrier densities as a function of dimensionless temperature $T/T_{\text{F}}$, where $T_{\text{F}}=E_{\text{F}}/k_{\text{B}}$ is the Fermi temperature ($E_{\text{F}}=\hbar\upsilon_{\text{F}}\sqrt{\pi n}$), Fig.~\ref{fig:concept}(f), and used the alternative form of Mott's equation, which relates the temperature-dependent thermopower and $T_{\text{F}}$, with $T\ll T_{\text{F}}$ \cite{hwang_theory_2009}:
\begin{equation} 
S(T) = -\frac{\pi^2 k_{\text{B}}}{3e}\frac{T}{T_{\text{F}}}\left(m+1\right),
\label{eq:TF_Mott}
\end{equation}
where the parameter $m$ represents the power-law energy-dependent scattering time $\tau(E)\propto E^{m}$. In monolayer graphene at high carrier densities ($n>10^{12}$~cm\textsuperscript{-2}) the dominant scattering is by the screened Coulomb impurities \cite{hwang_carrier_2007,hwang_theory_2009,dean_boron_2010} and the scattering time is energy-independent ($m=0$). In this case $S(T)$ is linear with temperature at a given $T_{\text{F}}$. At lower densities the scattering from the unscreened impurities ($m=1$) becomes more important and results in a superlinear $S(T)$ \cite{hwang_theory_2009,duan_high_2016}.

On the contrary, for our device the $S(T)$ values measured at low density of $n=\pm 2.2 \times 10^{11}$~cm\textsuperscript{-2} (red and blue stars in Fig.~\ref{fig:concept}(f)) were better described as being due to the screened impurities scattering ($m=0$). This unexpected result can possibly be explained by the presence of the relatively large density of charge fluctuations ($\delta n \sim 2\times 10^{11}$~cm\textsuperscript{-2}), which facilitates efficient screening.

At higher concentrations the cloned Dirac cones and the associated van Hove singularities start to play a more prominent role in  the thermoelectric response. Apart from the sign reversals associated with the Dirac points  at $n=0, \pm n_{0}$, we also observed a gradual change of thermopower sign at around $\pm0.5n_{0}$. This change is driven by the gradual increase of the carrier effective mass and eventual change in its sign at the van Hove singularity. Note at finite temperature the thermopower does not change the sign precisely at the singularity, but it happens in its vicinity due to the different relative contribution to the total thermopower from the carriers with different sign of effective mass. The competing contributions from different types of charge carriers is illustrated in Fig.~\ref{fig:concept}(f), where the thermopower at $n=\pm0.8n_{0}$ (red and blue triangles) is presented.  Contrary to the linear increase of $S(T/T_{\text{F}})$ seen near the mNP, the thermopower in the vicinity of the van Hove singularity is strongly non-monotonic with temperature. Due to thermal excitation, the carriers with the opposite sign of the effective mass become more involved in the thermometric transport, compensating for the initial low-temperature thermopower value. This behaviour is more pronounced for the eSNP, and can be explained by the smaller energy bandwidth of the secondary Dirac spectrum, which was reported to be one-third of that for the hSNP ($\sim25$~meV versus $\sim75$~meV) \cite{Yu2014}. Thus the eSNP-related features are smeared faster by the temperature with the overall thermoelectric response approaching the high-$T$ value for unaligned graphene given by Eq.~\ref{eq:TF_Mott} with $\upsilon_{\text{F}}=10^{6}$~m/s, $T=210$~K and $m=0$ (dashed red curves in Fig.~\ref{fig:concept}(d)). 

We also assessed the thermopower temperature dependence at higher carrier concentration $n=\pm1.6n_{0}$, away from SNPs and the related van Hove singularities (red and blue circles in Fig.~\ref{fig:concept}(f)). Surprisingly, the observed $S(T/T_{\text{F}})$ trend is well described by Eq.~\ref{eq:TF_Mott} if the parameter $m=1$, which is supposed to correspond to the unscreened Coulomb impurities (dashed lines in Fig.~\ref{fig:concept}(f)) and seems unlikely at such high carrier concentration ($n=3.52\times10^{12}$~cm\textsuperscript{12}). As discussed above, the transport at $|n|>1.2n_{0}$ is governed by the superlattice-induced Umklapp electron-electron scattering manifested in the quadratic temperature dependence of the excess resistivity $\Delta\rho$, shown in Fig.~\ref{fig:concept}(b). Potentially the Umklapp processes are responsible for this unexpected $S(T/T_{\text{F}})$ enhancement, however, further experimental and theoretical investigations are needed to confirm that.

\begin{figure*}[tbp]
	\includegraphics*[angle=0, trim=0mm 0mm 0mm 0mm, width=1\textwidth]{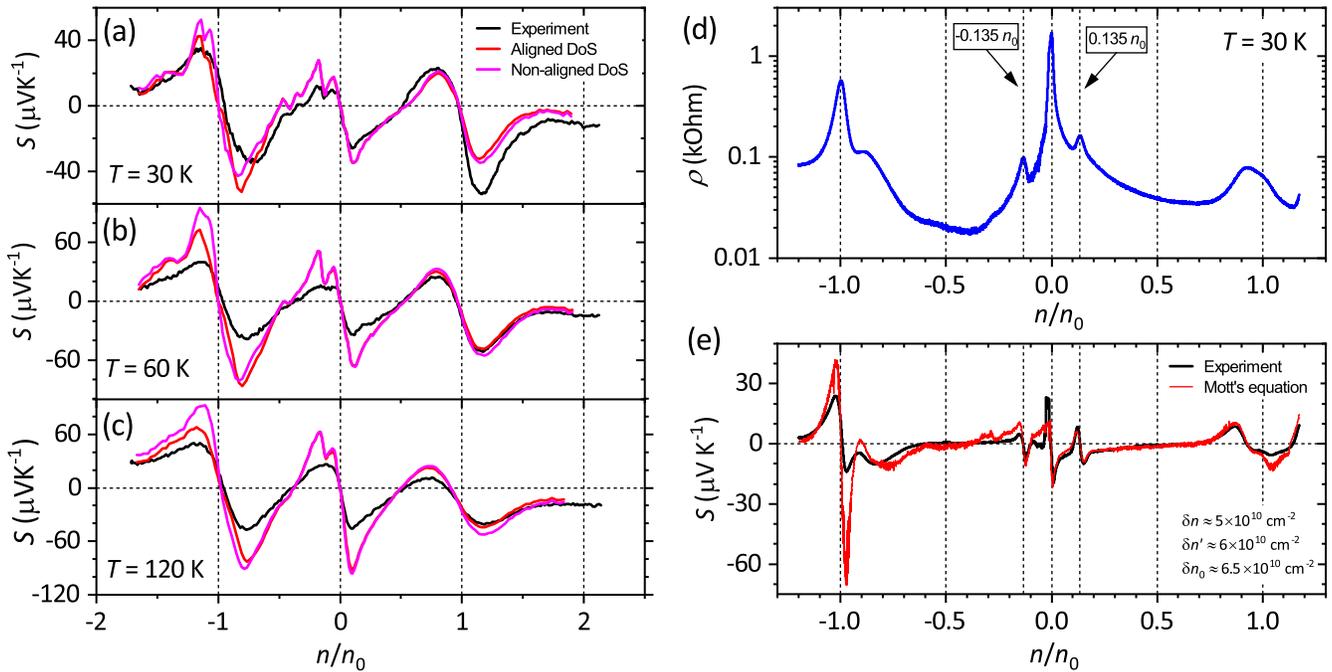}
	\caption{
		\label{fig:SmeasurementsandMott}
		\textbf{Thermoelectric measurements and approximation with Mott's equation.}
		(a-c) The measured thermopower, ${S}$ as a function of the normalised carrier density, ${n/n_0}$, at three device temperatures, ${T}$, (black curves) compared with the model based on Mott's equation in Eq.~\ref{eq:DifferentialMott} with a "non-aligned" DoS  (magenta curves) and "aligned" DoS (red curves). (d) Resistivity, $\rho$, as a function of the normalised density, $n/n_{0}$, at $T=30$~K for the high-mobility double-aligned hBN/graphene/hBN sample. The arrows point out the charge density at which the additional super-moir\'{e} Dirac peaks appear ($n' =\pm0.135n_{0}$, where $n_{0}=4.58\times10^{12}$~cm\textsuperscript{2}). (e) The measured thermopower, ${S}$ as a function of the normalised carrier density, ${n/n_0}$, at ${T}=30$~K, (black curve) for the double-aligned hBN/graphene/hBN sample compared with the predicted value given by Mott's equation in Eq.~\ref{eq:DifferentialMott} with a "non-aligned" DoS  (red curve) with the corresponding carrier density uncertainties $\delta n$, $\delta n'$ and $\delta n_{0}$.}
\end{figure*}

Another frequently used approach to the thermopower data analysis is the direct application of the differential form of Mott's equation, Eq.~\ref{eq:DifferentialMott}, with the derivative $\nicefrac{\partial \sigma}{\partial E}$ expressed as $\nicefrac{\partial \sigma}{\partial E} = \nicefrac{\partial \sigma}{\partial n}\times D(E)$, where $D(E) = \nicefrac{\partial n}{\partial E}$ is the density of states (DoS) taken at $E = E_{\text{F}}$. 
This approach relies on the numerical differentiation of the experimental conductivity data to find $\nicefrac{\partial\sigma}{\partial n}$ at a given $E_{\text{F}}$, explicit knowledge of the density of states, $D(E)$ \cite{Ghahari2016,zuev_thermoelectric_2009,wei_anomalous_2009,Nam_bilayer_TE,Wang_bilayer_TE} and the assumption that the measured values of $\sigma(n)$ contain all of the information on scattering mechanisms in Eq.~\ref{eq:DifferentialMott}.

For our aligned graphene/hBN sample we tested the above approach to approximate the experimental dependence of the thermopower on carrier concentration. Fig.~\ref{fig:SmeasurementsandMott}(a-c) shows the measured $S(n/n_{0})$ at three temperatures (black curves) along with the prediction produced with Eq.~\ref{eq:DifferentialMott} (red and magenta curves). The experimental values of the conductivity and its derivative were extracted from the transport data presented in Fig.~\ref{fig:concept}(a). The DoS has been modelled in two different ways. For the first one, all spectral features related to the moir\'{e} superlattice were ignored and the density of states for a non-aligned monolayer graphene was used, $D(E)=2E/(\pi\hbar^2\upsilon_{\text{F}}^2)$ (magenta curve). For the other one, the model DoS of an aligned graphene was adapted from the experimental data by Yu \textit{et al.} \cite{Yu2014} obtained in their quantum capacitance measurements (red curve). As can be seen, both ways provide a qualitative approximation of the experimental $S(n/n_{0})$ values incorporating all of the main peculiarities associated with the presence of the main and secondary Dirac points. Note the minor difference between the predicted values for the "aligned" and "non-aligned" DoS (red and magenta curves in Fig.~\ref{fig:SmeasurementsandMott}(a-c)), which indicates the relatively small role played by the differences in DoS. A somewhat better quantitative agreement is achieved for positive $n/n_{0}$ when the resistance changes smoothly with the carrier density. The major discrepancies appear near the regions of rapid resistance change (close to mNP and hSNP).

To further evaluate the applicability of the differential Mott's equation (Eq.~\ref{eq:DifferentialMott}) we measured the thermoelectric response from a high-quality hBN/graphene/hBN system at low temperature (carrier mobility $\sim4.5\times10^{5}$~cm\textsuperscript{2}V\textsuperscript{-1}s\textsuperscript{-1}), where the graphene monolayer is aligned to both top and bottom hBN crystals \cite{wang_new_generation_2019,wang_composite_2019,sun_correlated_2021}. The earlier report by Wang \textit{et} \textit{al.} \cite{wang_composite_2019} demonstrated that the double alignment of the graphene leads to the differential super-moir\'{e} reconstruction of the electron spectrum. The distinctive features of the super-moir\'{e} are its large period, unattainable in conventional single alignment structures, and additional secondary Dirac peaks in the vicinity of the mNP. Figure~\ref{fig:SmeasurementsandMott}(d) shows the transport data from our double-aligned structure. Apart from the conventional secondary Dirac peaks at $n=\pm n_{0}=\pm4.58\times10^{12}$~cm\textsuperscript{-2} (moir\'{e} period $\sim$10~nm, misalignment angle $\sim$0.98\textdegree), there are two additional secondary Dirac peaks at $n=\pm n'=0.135n_{0}$. The position of the latter implies the period of the moir\'{e} superlattice to be around 27.3~nm, which is significantly larger than the maximal 13.8~nm of an ideally aligned graphene/hBN and cannot be attributed to strain. We attribute this long period to the differential super-moir\'{e} caused by the the double alignment with top and bottom hBN crystals. Most likely, this super-moir\'{e} lattice is geometrically equivalent to the moir\'{e} lattice between two misaligned hBN crystals (angle $\sim 0.43^{\circ}$) \cite{wang_composite_2019}. Unfortunately, the finite dielectric strength of the SiO\textsubscript{2} backgate insulator did not allow to induce sufficient carrier density and observe the secondary Dirac peak at $n\approx1.5n_{0}$. This peak is expected from the second graphene-hBN misalignment angle estimated to be at around 1.41\textdegree. 

Expectedly, the thermoelectric response as a function of charge density exhibits a sign reversal for every Dirac point, including those attributed to the super-moir\'{e} reconstruction, Fig.~\ref{fig:SmeasurementsandMott}(e). Due to the better charge homogeneity the sign reversal regions are significantly narrower if compared to Fig.~\ref{fig:concept}(d), the peaks in thermopower are sharper and the small features are well resolved. The mNP sign reversal gives the charge inhomogeneity $\delta n\approx 5\times10^{10}$~cm\textsuperscript{-2} typical for high-quality graphene devices \cite{dean_boron_2010, kretinin_electronic_2014}. The additional super-moir\'{e} sign reversals are only marginally wider at $\delta n^{'} \approx 6\times10^{11}$~cm\textsuperscript{-2}, pointing to the differential reconstruction being less affected by the variation in misalignment angle or strain. Finally, the eSNP sign reversal gave the value of the filling concentration uncertainty $\delta n_{0} \approx 6.5\times10^{10}$~cm\textsuperscript{-2}, possibly also resulted from a small variation of the built-in strain ($\sim0.025\%$) introduced during the device fabrication.

The thermopower prediction, Fig.~\ref{fig:SmeasurementsandMott}(e), obtained from Eq.~\ref{eq:DifferentialMott} using the "non-aligned" DoS and numerical differentiation of the transport data shown in Fig.~\ref{fig:SmeasurementsandMott}(d), demonstrated reasonable qualitative agreement with the measurements, especially for the electron part of the spectrum. However, there is a substantial discrepancy with the experiment in the peak values of the thermopower near hSNP.

We attribute the discrepancies between Eq.~\ref{eq:DifferentialMott} and data in Fig.~\ref{fig:SmeasurementsandMott} observed for both samples to the inaccuracies of numerical differentiation of $\sigma$ rather than to any fundamental physical reasons. The discussed application of the differential Mott's equation is based on the assumption that the conductivity is the function composition of carrier density and energy $\sigma(n(E))$ or, equivalently, $\sigma(V_{\text{g}}(E))$. As a result, the derivative $\nicefrac{\partial \sigma}{\partial n}$ does not fully capture the effect of the energy-dependent scattering rate and leads to the discrepancies between the observed and predicted values of $S$ in the regions where this energy dependence is more significant.

\begin{figure}[tbp]
	\includegraphics*[angle=0, trim=0mm 0mm 0mm 0mm, width=1\linewidth]{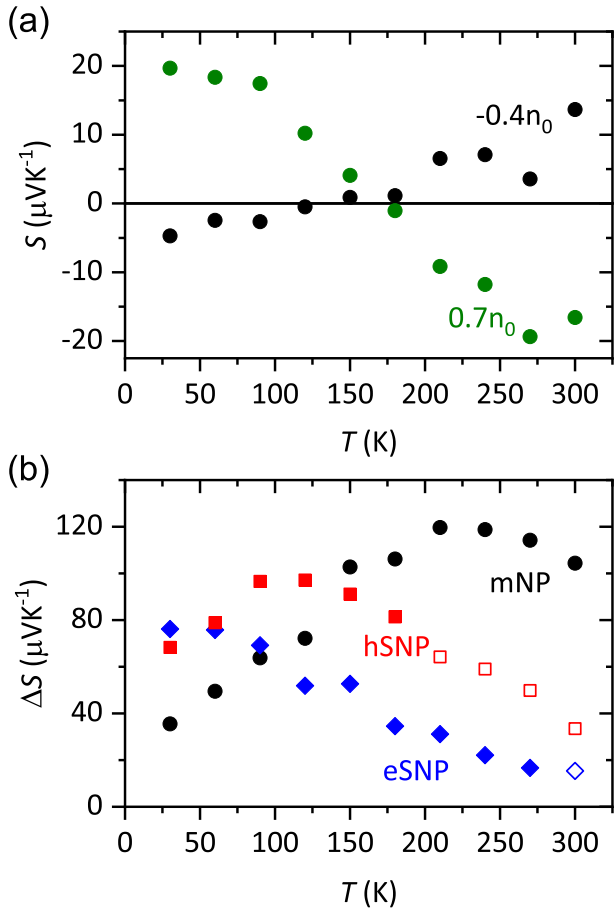}
	\caption{
		\label{fig:DeltaS}
		\textbf{Temperature dependence of thermopower reversal and of thermopower modulation at each neutrality point.}
		(a) Thermopower, ${S}$, as a function of temperature at two fixed carrier densities commensurate with the holes regime zero crossing point (${-0.4n_0}$) and electron regime zero crossing point (${0.7n_0}$). We normalise for $n_0$ at the SNP (2.2 \texttimes 10\textsuperscript{12} cm\textsuperscript{-2}). (b) The difference between the respective local maximum and minimum thermopower near each neutrality point, $\Delta S$, as a function of temperature. The open symbols represent estimates where the measurement could not extend the carrier density to the maximum (hSNP) and minimum (eSNP) due to risk of damaging the device by gating at higher global temperature, so the values shown are those obtained from the highest carrier density measurement achieved.}
\end{figure}

At intermediate carrier density ${n \approx\pm 0.5 n_0}$, in the regions between the mNP and the SNPs, the thermopower exhibits a temperature-driven sign change. 
Such a reversal is shown for our single-aligned graphene/hBN superlattice in Fig.~\ref{fig:DeltaS}(a). Here we have tuned the carrier density within each side of the mNP to achieve the desired reversals at a temperature of about 150\,K. 
Notably, both polarity reversals (negative to positive thermopower and viceversa) are achievable by the same device. 
Consequently, this polarity reversal of $S$, as a function of temperature, can be engineered and may be of technological significance. Note that a temperature-driven reversal is also exhibited in some common semi- and metallic thermoelectric materials (e.g. Bismuth \cite{Issi1979}, Platinum and Palladium \cite{Cusack1958}), where nonetheless it is caused by phonon drag \cite{Issi1979} and thus typically limited to low temperatures and small values of $S$. In our graphene/hBN superlattice a more robust effect is observed, across a wide temperature range, which is enabled by the vicinity of the van Hove singularity and the competition of thermally excited carriers with opposite sign of the effective mass.

To further elucidate the role of the different neutrality points on the temperature evolution of the thermopower, we subtracted the values of the maxima and minima of the thermopower in the vicinity of each neutrality point. The resulting thermopower modulation, ${\Delta S}$, is shown in Fig. \ref{fig:DeltaS}(b). Contrary to the predictions from Mott's equation, we observe a clear non-monotonic temperature dependence of this thermopower modulation across each neutrality point. In principle, a deviation from Mott's equation is expected at higher device temperature, as the condition $T \ll T_F$ is no longer valid \cite{Nam2010}. Nevertheless, we see a markedly distinct behaviour with ${\Delta S}$ achieving a maximum at low temperature for the eSNP, whereas for the hSNP a maximum is achieved at an intermediate temperature of 120\,K. For the main NP we observe a weak saturation at high temperature. We relate the observed hierarchy with the distinct DoS at the different neutrality points, i.e.\ degeneracy and Fermi velocity as discussed in the context of Eq.~\ref{eq:Delta}. The smaller energy bandwidth of the eSNP leads to a faster temperature smearing of its ${\Delta S}$ signature, with a correspondingly more robust response for the hSNP.

\section*{Conclusion}

We observed that there are additional thermopower features which are commensurate with secondary charge neutrality peaks and conform to the profile given by the differential form of Mott's equation. The discrepancies in the thermopower magnitudes between measurement and equation, at the mNP and hSNP, reveal a limitation in the applicability of the equation. The measurements have enabled us to quantify the charge density fluctuations, the role of thermally excited intrinsic carriers, and the strain placed on the graphene by the hBN, which in-turn has hinted at the presence of an obscured band gap. We have found that the temperature evolution of the thermopower reversal region associated with the NPs offers information on the degeneracies related to the DoS in each NP. This investigation has also enabled an insight into the temperature dependence of scattering mechanisms in an aligned device. Here we found the signature of carrier scattering by screened impurities at low carrier density; contrary to that of non-aligned graphene. At high carrier density the thermopower indicates a possible role of superlattice-induced Umklapp scattering. Consequently, thermoelectric measurements may be used as a sensitive probe of the electronic spectrum of a graphene-hBN superlattice and hBN-graphene-hBN super-moir\'{e} lattice. Furthermore, it is possible to control the thermoelectric properties of graphene by engineering its electronic structure. An interesting feature of this is the ability to select the carrier density where a thermopower reversal occurs at a certain temperature. This may be of technological significance, where the tunability of the temperature at which cooling is changed to heating may be used to regulate temperature in thermoelectric applications.

\begin{acknowledgments}
This work is supported by the UK EPSRC via a PhD studentship and Doctoral Prize Fellowship. J.H.E. acknowledges the support of the Materials Engineering and Processing program of the National Science Foundation, Award Number CMMI 1538127, A.V.K. acknowledges support from EPSRC (grant EP/V036343/1) and from the European Graphene Flagship Project, R.G. acknowledges support from Royal Society, ERC Consolidator grant QTWIST (101001515) and EPSRC grant numbers EP/V007033/1, EP/S030719/1 and EP/V026496/1
\end{acknowledgments}


\begin{thebibliography}{45}%
\makeatletter
\providecommand \@ifxundefined [1]{%
 \@ifx{#1\undefined}
}%
\providecommand \@ifnum [1]{%
 \ifnum #1\expandafter \@firstoftwo
 \else \expandafter \@secondoftwo
 \fi
}%
\providecommand \@ifx [1]{%
 \ifx #1\expandafter \@firstoftwo
 \else \expandafter \@secondoftwo
 \fi
}%
\providecommand \natexlab [1]{#1}%
\providecommand \enquote  [1]{``#1''}%
\providecommand \bibnamefont  [1]{#1}%
\providecommand \bibfnamefont [1]{#1}%
\providecommand \citenamefont [1]{#1}%
\providecommand \href@noop [0]{\@secondoftwo}%
\providecommand \href [0]{\begingroup \@sanitize@url \@href}%
\providecommand \@href[1]{\@@startlink{#1}\@@href}%
\providecommand \@@href[1]{\endgroup#1\@@endlink}%
\providecommand \@sanitize@url [0]{\catcode `\\12\catcode `\$12\catcode
  `\&12\catcode `\#12\catcode `\^12\catcode `\_12\catcode `\%12\relax}%
\providecommand \@@startlink[1]{}%
\providecommand \@@endlink[0]{}%
\providecommand \url  [0]{\begingroup\@sanitize@url \@url }%
\providecommand \@url [1]{\endgroup\@href {#1}{\urlprefix }}%
\providecommand \urlprefix  [0]{URL }%
\providecommand \Eprint [0]{\href }%
\providecommand \doibase [0]{https://doi.org/}%
\providecommand \selectlanguage [0]{\@gobble}%
\providecommand \bibinfo  [0]{\@secondoftwo}%
\providecommand \bibfield  [0]{\@secondoftwo}%
\providecommand \translation [1]{[#1]}%
\providecommand \BibitemOpen [0]{}%
\providecommand \bibitemStop [0]{}%
\providecommand \bibitemNoStop [0]{.\EOS\space}%
\providecommand \EOS [0]{\spacefactor3000\relax}%
\providecommand \BibitemShut  [1]{\csname bibitem#1\endcsname}%
\let\auto@bib@innerbib\@empty
\bibitem [{\citenamefont {He}\ and\ \citenamefont
  {Tritt}(2017)}]{He_Tritt_review}%
  \BibitemOpen
  \bibfield  {author} {\bibinfo {author} {\bibfnamefont {J.}~\bibnamefont
  {He}}\ and\ \bibinfo {author} {\bibfnamefont {T.~M.}\ \bibnamefont {Tritt}},\
  }\bibfield  {title} {\bibinfo {title} {Advances in thermoelectric materials
  research: Looking back and moving forward},\ }\href
  {https://doi.org/10.1126/science.aak9997} {\bibfield  {journal} {\bibinfo
  {journal} {Science}\ }\textbf {\bibinfo {volume} {357}},\ \bibinfo {pages}
  {eaak9997} (\bibinfo {year} {2017})}\BibitemShut {NoStop}%
\bibitem [{\citenamefont {van Houten}\ \emph {et~al.}(1992)\citenamefont {van
  Houten}, \citenamefont {Molenkamp}, \citenamefont {Beenakker},\ and\
  \citenamefont {Foxon}}]{Houtent1992}%
  \BibitemOpen
  \bibfield  {author} {\bibinfo {author} {\bibfnamefont {H.}~\bibnamefont {van
  Houten}}, \bibinfo {author} {\bibfnamefont {L.~W.}\ \bibnamefont
  {Molenkamp}}, \bibinfo {author} {\bibfnamefont {C.~W.~J.}\ \bibnamefont
  {Beenakker}},\ and\ \bibinfo {author} {\bibfnamefont {C.~T.}\ \bibnamefont
  {Foxon}},\ }\bibfield  {title} {\bibinfo {title} {{Thermo-electric properties
  of quantum point contacts}},\ }\href
  {https://doi.org/10.1088/0268-1242/7/3B/052} {\bibfield  {journal} {\bibinfo
  {journal} {Semiconductor Science and Technology}\ }\textbf {\bibinfo {volume}
  {7}},\ \bibinfo {pages} {B215} (\bibinfo {year} {1992})}\BibitemShut
  {NoStop}%
\bibitem [{\citenamefont {Ghahari}\ \emph
  {et~al.}(2016{\natexlab{a}})\citenamefont {Ghahari}, \citenamefont {Xie},
  \citenamefont {Taniguchi}, \citenamefont {Watanabe}, \citenamefont {Foster},\
  and\ \citenamefont {Kim}}]{Ghahari2016}%
  \BibitemOpen
  \bibfield  {author} {\bibinfo {author} {\bibfnamefont {F.}~\bibnamefont
  {Ghahari}}, \bibinfo {author} {\bibfnamefont {H.~Y.}\ \bibnamefont {Xie}},
  \bibinfo {author} {\bibfnamefont {T.}~\bibnamefont {Taniguchi}}, \bibinfo
  {author} {\bibfnamefont {K.}~\bibnamefont {Watanabe}}, \bibinfo {author}
  {\bibfnamefont {M.~S.}\ \bibnamefont {Foster}},\ and\ \bibinfo {author}
  {\bibfnamefont {P.}~\bibnamefont {Kim}},\ }\bibfield  {title} {\bibinfo
  {title} {{Enhanced Thermoelectric Power in Graphene: Violation of the Mott
  Relation by Inelastic Scattering}},\ }\href
  {https://doi.org/10.1103/PhysRevLett.116.136802} {\bibfield  {journal}
  {\bibinfo  {journal} {Physical Review Letters}\ }\textbf {\bibinfo {volume}
  {116}},\ \bibinfo {pages} {136802} (\bibinfo {year} {2016}{\natexlab{a}})},\
  \Eprint {https://arxiv.org/abs/1601.05859} {arXiv:1601.05859} \BibitemShut
  {NoStop}%
\bibitem [{\citenamefont {Mahapatra}\ \emph {et~al.}(2020)\citenamefont
  {Mahapatra}, \citenamefont {Ghawri}, \citenamefont {Garg}, \citenamefont
  {Mandal}, \citenamefont {Watanabe}, \citenamefont {Taniguchi}, \citenamefont
  {Jain}, \citenamefont {Mukerjee},\ and\ \citenamefont
  {Ghosh}}]{Mahapatra2020}%
  \BibitemOpen
  \bibfield  {author} {\bibinfo {author} {\bibfnamefont {P.~S.}\ \bibnamefont
  {Mahapatra}}, \bibinfo {author} {\bibfnamefont {B.}~\bibnamefont {Ghawri}},
  \bibinfo {author} {\bibfnamefont {M.}~\bibnamefont {Garg}}, \bibinfo {author}
  {\bibfnamefont {S.}~\bibnamefont {Mandal}}, \bibinfo {author} {\bibfnamefont
  {K.}~\bibnamefont {Watanabe}}, \bibinfo {author} {\bibfnamefont
  {T.}~\bibnamefont {Taniguchi}}, \bibinfo {author} {\bibfnamefont
  {M.}~\bibnamefont {Jain}}, \bibinfo {author} {\bibfnamefont {S.}~\bibnamefont
  {Mukerjee}},\ and\ \bibinfo {author} {\bibfnamefont {A.}~\bibnamefont
  {Ghosh}},\ }\bibfield  {title} {\bibinfo {title} {{Misorientation-Controlled
  Cross-Plane Thermoelectricity in Twisted Bilayer Graphene}},\ }\href
  {https://doi.org/10.1103/PhysRevLett.125.226802} {\bibfield  {journal}
  {\bibinfo  {journal} {Physical Review Letters}\ }\textbf {\bibinfo {volume}
  {125}},\ \bibinfo {pages} {226802} (\bibinfo {year} {2020})}\BibitemShut
  {NoStop}%
\bibitem [{\citenamefont {Paul}\ \emph {et~al.}(2022)\citenamefont {Paul},
  \citenamefont {Ghosh}, \citenamefont {Chakraborty}, \citenamefont {Roy},
  \citenamefont {Dutta}, \citenamefont {Watanabe}, \citenamefont {Taniguchi},
  \citenamefont {Panda}, \citenamefont {Agarwala}, \citenamefont {Mukerjee},
  \citenamefont {Banerjee},\ and\ \citenamefont {Das}}]{Paul2022}%
  \BibitemOpen
  \bibfield  {author} {\bibinfo {author} {\bibfnamefont {A.~K.}\ \bibnamefont
  {Paul}}, \bibinfo {author} {\bibfnamefont {A.}~\bibnamefont {Ghosh}},
  \bibinfo {author} {\bibfnamefont {S.}~\bibnamefont {Chakraborty}}, \bibinfo
  {author} {\bibfnamefont {U.}~\bibnamefont {Roy}}, \bibinfo {author}
  {\bibfnamefont {R.}~\bibnamefont {Dutta}}, \bibinfo {author} {\bibfnamefont
  {K.}~\bibnamefont {Watanabe}}, \bibinfo {author} {\bibfnamefont
  {T.}~\bibnamefont {Taniguchi}}, \bibinfo {author} {\bibfnamefont
  {A.}~\bibnamefont {Panda}}, \bibinfo {author} {\bibfnamefont
  {A.}~\bibnamefont {Agarwala}}, \bibinfo {author} {\bibfnamefont
  {S.}~\bibnamefont {Mukerjee}}, \bibinfo {author} {\bibfnamefont
  {S.}~\bibnamefont {Banerjee}},\ and\ \bibinfo {author} {\bibfnamefont
  {A.}~\bibnamefont {Das}},\ }\bibfield  {title} {\bibinfo {title}
  {{Interaction-driven giant thermopower in magic-angle twisted bilayer
  graphene}},\ }\bibfield  {journal} {\bibinfo  {journal} {Nature Physics}\
  }\href {https://doi.org/10.1038/s41567-022-01574-3}
  {10.1038/s41567-022-01574-3} (\bibinfo {year} {2022})\BibitemShut {NoStop}%
\bibitem [{\citenamefont {Ghawri}\ \emph {et~al.}(2022)\citenamefont {Ghawri},
  \citenamefont {Mahapatra}, \citenamefont {Garg}, \citenamefont {Mandal},
  \citenamefont {Bhowmik}, \citenamefont {Jayaraman}, \citenamefont {Soni},
  \citenamefont {Watanabe}, \citenamefont {Taniguchi}, \citenamefont
  {Krishnamurthy}, \citenamefont {Jain}, \citenamefont {Banerjee},
  \citenamefont {Chandni},\ and\ \citenamefont {Ghosh}}]{Ghawri2022}%
  \BibitemOpen
  \bibfield  {author} {\bibinfo {author} {\bibfnamefont {B.}~\bibnamefont
  {Ghawri}}, \bibinfo {author} {\bibfnamefont {P.~S.}\ \bibnamefont
  {Mahapatra}}, \bibinfo {author} {\bibfnamefont {M.}~\bibnamefont {Garg}},
  \bibinfo {author} {\bibfnamefont {S.}~\bibnamefont {Mandal}}, \bibinfo
  {author} {\bibfnamefont {S.}~\bibnamefont {Bhowmik}}, \bibinfo {author}
  {\bibfnamefont {A.}~\bibnamefont {Jayaraman}}, \bibinfo {author}
  {\bibfnamefont {R.}~\bibnamefont {Soni}}, \bibinfo {author} {\bibfnamefont
  {K.}~\bibnamefont {Watanabe}}, \bibinfo {author} {\bibfnamefont
  {T.}~\bibnamefont {Taniguchi}}, \bibinfo {author} {\bibfnamefont {H.~R.}\
  \bibnamefont {Krishnamurthy}}, \bibinfo {author} {\bibfnamefont
  {M.}~\bibnamefont {Jain}}, \bibinfo {author} {\bibfnamefont {S.}~\bibnamefont
  {Banerjee}}, \bibinfo {author} {\bibfnamefont {U.}~\bibnamefont {Chandni}},\
  and\ \bibinfo {author} {\bibfnamefont {A.}~\bibnamefont {Ghosh}},\ }\bibfield
   {title} {\bibinfo {title} {{Breakdown of semiclassical description of
  thermoelectricity in near-magic angle twisted bilayer graphene}},\ }\href
  {https://doi.org/10.1038/s41467-022-29198-4} {\bibfield  {journal} {\bibinfo
  {journal} {Nature Communications}\ }\textbf {\bibinfo {volume} {13}},\
  \bibinfo {pages} {1522} (\bibinfo {year} {2022})}\BibitemShut {NoStop}%
\bibitem [{\citenamefont {Xu}\ \emph {et~al.}(2014)\citenamefont {Xu},
  \citenamefont {Li},\ and\ \citenamefont
  {Duan}}]{Graphene_TE_review_Xu_Li_Duan}%
  \BibitemOpen
  \bibfield  {author} {\bibinfo {author} {\bibfnamefont {Y.}~\bibnamefont
  {Xu}}, \bibinfo {author} {\bibfnamefont {Z.}~\bibnamefont {Li}},\ and\
  \bibinfo {author} {\bibfnamefont {W.}~\bibnamefont {Duan}},\ }\bibfield
  {title} {\bibinfo {title} {Thermal and thermoelectric properties of
  graphene},\ }\href {https://doi.org/https://doi.org/10.1002/smll.201303701}
  {\bibfield  {journal} {\bibinfo  {journal} {Small}\ }\textbf {\bibinfo
  {volume} {10}},\ \bibinfo {pages} {2182} (\bibinfo {year}
  {2014})}\BibitemShut {NoStop}%
\bibitem [{\citenamefont {Cutler}\ and\ \citenamefont
  {Mott}(1969)}]{cutler_observation_1969}%
  \BibitemOpen
  \bibfield  {author} {\bibinfo {author} {\bibfnamefont {M.}~\bibnamefont
  {Cutler}}\ and\ \bibinfo {author} {\bibfnamefont {N.~F.}\ \bibnamefont
  {Mott}},\ }\bibfield  {title} {\bibinfo {title} {Observation of {Anderson}
  {Localization} in an {Electron} {Gas}},\ }\href
  {https://doi.org/10.1103/PhysRev.181.1336} {\bibfield  {journal} {\bibinfo
  {journal} {Phys. Rev.}\ }\textbf {\bibinfo {volume} {181}},\ \bibinfo {pages}
  {1336} (\bibinfo {year} {1969})}\BibitemShut {NoStop}%
\bibitem [{\citenamefont {Zuev}\ \emph {et~al.}(2009)\citenamefont {Zuev},
  \citenamefont {Chang},\ and\ \citenamefont {Kim}}]{zuev_thermoelectric_2009}%
  \BibitemOpen
  \bibfield  {author} {\bibinfo {author} {\bibfnamefont {Y.~M.}\ \bibnamefont
  {Zuev}}, \bibinfo {author} {\bibfnamefont {W.}~\bibnamefont {Chang}},\ and\
  \bibinfo {author} {\bibfnamefont {P.}~\bibnamefont {Kim}},\ }\bibfield
  {title} {\bibinfo {title} {Thermoelectric and {Magnetothermoelectric}
  {Transport} {Measurements} of {Graphene}},\ }\href
  {https://doi.org/10.1103/PhysRevLett.102.096807} {\bibfield  {journal}
  {\bibinfo  {journal} {Phys. Rev. Lett.}\ }\textbf {\bibinfo {volume} {102}},\
  \bibinfo {pages} {096807} (\bibinfo {year} {2009})}\BibitemShut {NoStop}%
\bibitem [{\citenamefont {Wei}\ \emph {et~al.}(2009)\citenamefont {Wei},
  \citenamefont {Bao}, \citenamefont {Pu}, \citenamefont {Lau},\ and\
  \citenamefont {Shi}}]{wei_anomalous_2009}%
  \BibitemOpen
  \bibfield  {author} {\bibinfo {author} {\bibfnamefont {P.}~\bibnamefont
  {Wei}}, \bibinfo {author} {\bibfnamefont {W.}~\bibnamefont {Bao}}, \bibinfo
  {author} {\bibfnamefont {Y.}~\bibnamefont {Pu}}, \bibinfo {author}
  {\bibfnamefont {C.~N.}\ \bibnamefont {Lau}},\ and\ \bibinfo {author}
  {\bibfnamefont {J.}~\bibnamefont {Shi}},\ }\bibfield  {title} {\bibinfo
  {title} {Anomalous {Thermoelectric} {Transport} of {Dirac} {Particles} in
  {Graphene}},\ }\href {https://doi.org/10.1103/PhysRevLett.102.166808}
  {\bibfield  {journal} {\bibinfo  {journal} {Phys. Rev. Lett.}\ }\textbf
  {\bibinfo {volume} {102}},\ \bibinfo {pages} {166808} (\bibinfo {year}
  {2009})}\BibitemShut {NoStop}%
\bibitem [{\citenamefont {Checkelsky}\ and\ \citenamefont
  {Ong}(2009)}]{checkelsky_thermopower_2009}%
  \BibitemOpen
  \bibfield  {author} {\bibinfo {author} {\bibfnamefont {J.~G.}\ \bibnamefont
  {Checkelsky}}\ and\ \bibinfo {author} {\bibfnamefont {N.~P.}\ \bibnamefont
  {Ong}},\ }\bibfield  {title} {\bibinfo {title} {Thermopower and {Nernst}
  effect in graphene in a magnetic field},\ }\href
  {https://doi.org/10.1103/PhysRevB.80.081413} {\bibfield  {journal} {\bibinfo
  {journal} {Phys. Rev. B}\ }\textbf {\bibinfo {volume} {80}},\ \bibinfo
  {pages} {081413} (\bibinfo {year} {2009})}\BibitemShut {NoStop}%
\bibitem [{\citenamefont {Duan}\ \emph {et~al.}(2016)\citenamefont {Duan},
  \citenamefont {Wang}, \citenamefont {Lai}, \citenamefont {Li}, \citenamefont
  {Watanabe}, \citenamefont {Taniguchi}, \citenamefont {Zebarjadi},\ and\
  \citenamefont {Andrei}}]{duan_high_2016}%
  \BibitemOpen
  \bibfield  {author} {\bibinfo {author} {\bibfnamefont {J.}~\bibnamefont
  {Duan}}, \bibinfo {author} {\bibfnamefont {X.}~\bibnamefont {Wang}}, \bibinfo
  {author} {\bibfnamefont {X.}~\bibnamefont {Lai}}, \bibinfo {author}
  {\bibfnamefont {G.}~\bibnamefont {Li}}, \bibinfo {author} {\bibfnamefont
  {K.}~\bibnamefont {Watanabe}}, \bibinfo {author} {\bibfnamefont
  {T.}~\bibnamefont {Taniguchi}}, \bibinfo {author} {\bibfnamefont
  {M.}~\bibnamefont {Zebarjadi}},\ and\ \bibinfo {author} {\bibfnamefont
  {E.~Y.}\ \bibnamefont {Andrei}},\ }\bibfield  {title} {\bibinfo {title} {High
  thermoelectricpower factor in graphene/{hBN} devices},\ }\href
  {https://doi.org/10.1073/pnas.1615913113} {\bibfield  {journal} {\bibinfo
  {journal} {PNAS}\ }\textbf {\bibinfo {volume} {113}},\ \bibinfo {pages}
  {14272} (\bibinfo {year} {2016})}\BibitemShut {NoStop}%
\bibitem [{\citenamefont {Nam}\ \emph {et~al.}(2010{\natexlab{a}})\citenamefont
  {Nam}, \citenamefont {Ki},\ and\ \citenamefont {Lee}}]{Nam_bilayer_TE}%
  \BibitemOpen
  \bibfield  {author} {\bibinfo {author} {\bibfnamefont {S.-G.}\ \bibnamefont
  {Nam}}, \bibinfo {author} {\bibfnamefont {D.-K.}\ \bibnamefont {Ki}},\ and\
  \bibinfo {author} {\bibfnamefont {H.-J.}\ \bibnamefont {Lee}},\ }\bibfield
  {title} {\bibinfo {title} {Thermoelectric transport of massive dirac fermions
  in bilayer graphene},\ }\href {https://doi.org/10.1103/PhysRevB.82.245416}
  {\bibfield  {journal} {\bibinfo  {journal} {Phys. Rev. B}\ }\textbf {\bibinfo
  {volume} {82}},\ \bibinfo {pages} {245416} (\bibinfo {year}
  {2010}{\natexlab{a}})}\BibitemShut {NoStop}%
\bibitem [{\citenamefont {Wang}\ \emph {et~al.}(2011)\citenamefont {Wang},
  \citenamefont {Lu}, \citenamefont {Hao}, \citenamefont {Lee}, \citenamefont
  {Lee}, \citenamefont {Lin}, \citenamefont {Cheng},\ and\ \citenamefont
  {Chen}}]{Wang_bilayer_TE}%
  \BibitemOpen
  \bibfield  {author} {\bibinfo {author} {\bibfnamefont {C.-R.}\ \bibnamefont
  {Wang}}, \bibinfo {author} {\bibfnamefont {W.-S.}\ \bibnamefont {Lu}},
  \bibinfo {author} {\bibfnamefont {L.}~\bibnamefont {Hao}}, \bibinfo {author}
  {\bibfnamefont {W.-L.}\ \bibnamefont {Lee}}, \bibinfo {author} {\bibfnamefont
  {T.-K.}\ \bibnamefont {Lee}}, \bibinfo {author} {\bibfnamefont
  {F.}~\bibnamefont {Lin}}, \bibinfo {author} {\bibfnamefont {I.-C.}\
  \bibnamefont {Cheng}},\ and\ \bibinfo {author} {\bibfnamefont {J.-Z.}\
  \bibnamefont {Chen}},\ }\bibfield  {title} {\bibinfo {title} {Enhanced
  thermoelectric power in dual-gated bilayer graphene},\ }\href
  {https://doi.org/10.1103/PhysRevLett.107.186602} {\bibfield  {journal}
  {\bibinfo  {journal} {Phys. Rev. Lett.}\ }\textbf {\bibinfo {volume} {107}},\
  \bibinfo {pages} {186602} (\bibinfo {year} {2011})}\BibitemShut {NoStop}%
\bibitem [{\citenamefont {Frisenda}\ \emph {et~al.}(2018)\citenamefont
  {Frisenda}, \citenamefont {Navarro-Moratalla}, \citenamefont {Gant},
  \citenamefont {Pérez De~Lara}, \citenamefont {Jarillo-Herrero},
  \citenamefont {Gorbachev},\ and\ \citenamefont
  {Castellanos-Gomez}}]{Frisenda_fab_review}%
  \BibitemOpen
  \bibfield  {author} {\bibinfo {author} {\bibfnamefont {R.}~\bibnamefont
  {Frisenda}}, \bibinfo {author} {\bibfnamefont {E.}~\bibnamefont
  {Navarro-Moratalla}}, \bibinfo {author} {\bibfnamefont {P.}~\bibnamefont
  {Gant}}, \bibinfo {author} {\bibfnamefont {D.}~\bibnamefont {Pérez
  De~Lara}}, \bibinfo {author} {\bibfnamefont {P.}~\bibnamefont
  {Jarillo-Herrero}}, \bibinfo {author} {\bibfnamefont {R.~V.}\ \bibnamefont
  {Gorbachev}},\ and\ \bibinfo {author} {\bibfnamefont {A.}~\bibnamefont
  {Castellanos-Gomez}},\ }\bibfield  {title} {\bibinfo {title} {Recent progress
  in the assembly of nanodevices and van der waals heterostructures by
  deterministic placement of 2d materials},\ }\href
  {https://doi.org/10.1039/C7CS00556C} {\bibfield  {journal} {\bibinfo
  {journal} {Chem. Soc. Rev.}\ }\textbf {\bibinfo {volume} {47}},\ \bibinfo
  {pages} {53} (\bibinfo {year} {2018})}\BibitemShut {NoStop}%
\bibitem [{\citenamefont {Woods}\ \emph {et~al.}(2014)\citenamefont {Woods},
  \citenamefont {Britnell}, \citenamefont {Eckmann}, \citenamefont {Ma},
  \citenamefont {Lu}, \citenamefont {Guo}, \citenamefont {Lin}, \citenamefont
  {Yu}, \citenamefont {Cao}, \citenamefont {Gorbachev}, \citenamefont
  {Kretinin}, \citenamefont {Park}, \citenamefont {Ponomarenko}, \citenamefont
  {Katsnelson}, \citenamefont {Gornostyrev}, \citenamefont {Watanabe},
  \citenamefont {Taniguchi}, \citenamefont {Casiraghi}, \citenamefont {Gao},
  \citenamefont {Geim},\ and\ \citenamefont
  {Novoselov}}]{Woods_commensurate_graphene_hBN}%
  \BibitemOpen
  \bibfield  {author} {\bibinfo {author} {\bibfnamefont {C.~R.}\ \bibnamefont
  {Woods}}, \bibinfo {author} {\bibfnamefont {L.}~\bibnamefont {Britnell}},
  \bibinfo {author} {\bibfnamefont {A.}~\bibnamefont {Eckmann}}, \bibinfo
  {author} {\bibfnamefont {R.~S.}\ \bibnamefont {Ma}}, \bibinfo {author}
  {\bibfnamefont {J.~C.}\ \bibnamefont {Lu}}, \bibinfo {author} {\bibfnamefont
  {H.~M.}\ \bibnamefont {Guo}}, \bibinfo {author} {\bibfnamefont
  {X.}~\bibnamefont {Lin}}, \bibinfo {author} {\bibfnamefont {G.~L.}\
  \bibnamefont {Yu}}, \bibinfo {author} {\bibfnamefont {Y.}~\bibnamefont
  {Cao}}, \bibinfo {author} {\bibfnamefont {R.~V.}\ \bibnamefont {Gorbachev}},
  \bibinfo {author} {\bibfnamefont {A.~V.}\ \bibnamefont {Kretinin}}, \bibinfo
  {author} {\bibfnamefont {J.}~\bibnamefont {Park}}, \bibinfo {author}
  {\bibfnamefont {L.~A.}\ \bibnamefont {Ponomarenko}}, \bibinfo {author}
  {\bibfnamefont {M.~I.}\ \bibnamefont {Katsnelson}}, \bibinfo {author}
  {\bibfnamefont {Y.~N.}\ \bibnamefont {Gornostyrev}}, \bibinfo {author}
  {\bibfnamefont {K.}~\bibnamefont {Watanabe}}, \bibinfo {author}
  {\bibfnamefont {T.}~\bibnamefont {Taniguchi}}, \bibinfo {author}
  {\bibfnamefont {C.}~\bibnamefont {Casiraghi}}, \bibinfo {author}
  {\bibfnamefont {H.-J.}\ \bibnamefont {Gao}}, \bibinfo {author} {\bibfnamefont
  {A.~K.}\ \bibnamefont {Geim}},\ and\ \bibinfo {author} {\bibfnamefont
  {K.~S.}\ \bibnamefont {Novoselov}},\ }\bibfield  {title} {\bibinfo {title}
  {Commensurate-incommensurate transition in graphene on hexagonal boron
  nitride},\ }\href {https://doi.org/10.1038/NPHYS2954} {\bibfield  {journal}
  {\bibinfo  {journal} {Nature Physics}\ }\textbf {\bibinfo {volume} {10}},\
  \bibinfo {pages} {451} (\bibinfo {year} {2014})}\BibitemShut {NoStop}%
\bibitem [{\citenamefont {Ponomarenko}\ \emph {et~al.}(2013)\citenamefont
  {Ponomarenko}, \citenamefont {Gorbachev}, \citenamefont {Yu}, \citenamefont
  {Elias}, \citenamefont {Jalil}, \citenamefont {Patel}, \citenamefont
  {Mishchenko}, \citenamefont {Mayorov}, \citenamefont {Woods}, \citenamefont
  {Wallbank}, \citenamefont {Mucha-Kruczynski}, \citenamefont {Piot},
  \citenamefont {Potemski}, \citenamefont {Grigorieva}, \citenamefont
  {Novoselov}, \citenamefont {Guinea}, \citenamefont {Fal’ko},\ and\
  \citenamefont {Geim}}]{ponomarenko_cloning_2013}%
  \BibitemOpen
  \bibfield  {author} {\bibinfo {author} {\bibfnamefont {L.~A.}\ \bibnamefont
  {Ponomarenko}}, \bibinfo {author} {\bibfnamefont {R.~V.}\ \bibnamefont
  {Gorbachev}}, \bibinfo {author} {\bibfnamefont {G.~L.}\ \bibnamefont {Yu}},
  \bibinfo {author} {\bibfnamefont {D.~C.}\ \bibnamefont {Elias}}, \bibinfo
  {author} {\bibfnamefont {R.}~\bibnamefont {Jalil}}, \bibinfo {author}
  {\bibfnamefont {A.~A.}\ \bibnamefont {Patel}}, \bibinfo {author}
  {\bibfnamefont {A.}~\bibnamefont {Mishchenko}}, \bibinfo {author}
  {\bibfnamefont {A.~S.}\ \bibnamefont {Mayorov}}, \bibinfo {author}
  {\bibfnamefont {C.~R.}\ \bibnamefont {Woods}}, \bibinfo {author}
  {\bibfnamefont {J.~R.}\ \bibnamefont {Wallbank}}, \bibinfo {author}
  {\bibfnamefont {M.}~\bibnamefont {Mucha-Kruczynski}}, \bibinfo {author}
  {\bibfnamefont {B.~A.}\ \bibnamefont {Piot}}, \bibinfo {author}
  {\bibfnamefont {M.}~\bibnamefont {Potemski}}, \bibinfo {author}
  {\bibfnamefont {I.~V.}\ \bibnamefont {Grigorieva}}, \bibinfo {author}
  {\bibfnamefont {K.~S.}\ \bibnamefont {Novoselov}}, \bibinfo {author}
  {\bibfnamefont {F.}~\bibnamefont {Guinea}}, \bibinfo {author} {\bibfnamefont
  {V.~I.}\ \bibnamefont {Fal’ko}},\ and\ \bibinfo {author} {\bibfnamefont
  {A.~K.}\ \bibnamefont {Geim}},\ }\bibfield  {title} {\bibinfo {title}
  {Cloning of {Dirac} fermions in graphene superlattices},\ }\href
  {https://doi.org/10.1038/nature12187} {\bibfield  {journal} {\bibinfo
  {journal} {Nature}\ }\textbf {\bibinfo {volume} {497}},\ \bibinfo {pages}
  {594} (\bibinfo {year} {2013})}\BibitemShut {NoStop}%
\bibitem [{\citenamefont {Hunt}\ \emph {et~al.}(2013)\citenamefont {Hunt},
  \citenamefont {Sanchez-Yamagishi}, \citenamefont {Young}, \citenamefont
  {Yankowitz}, \citenamefont {LeRoy}, \citenamefont {Watanabe}, \citenamefont
  {Taniguchi}, \citenamefont {Moon}, \citenamefont {Koshino}, \citenamefont
  {Jarillo-Herrero},\ and\ \citenamefont {Ashoori}}]{hunt_massive_2013}%
  \BibitemOpen
  \bibfield  {author} {\bibinfo {author} {\bibfnamefont {B.}~\bibnamefont
  {Hunt}}, \bibinfo {author} {\bibfnamefont {J.~D.}\ \bibnamefont
  {Sanchez-Yamagishi}}, \bibinfo {author} {\bibfnamefont {A.~F.}\ \bibnamefont
  {Young}}, \bibinfo {author} {\bibfnamefont {M.}~\bibnamefont {Yankowitz}},
  \bibinfo {author} {\bibfnamefont {B.~J.}\ \bibnamefont {LeRoy}}, \bibinfo
  {author} {\bibfnamefont {K.}~\bibnamefont {Watanabe}}, \bibinfo {author}
  {\bibfnamefont {T.}~\bibnamefont {Taniguchi}}, \bibinfo {author}
  {\bibfnamefont {P.}~\bibnamefont {Moon}}, \bibinfo {author} {\bibfnamefont
  {M.}~\bibnamefont {Koshino}}, \bibinfo {author} {\bibfnamefont
  {P.}~\bibnamefont {Jarillo-Herrero}},\ and\ \bibinfo {author} {\bibfnamefont
  {R.~C.}\ \bibnamefont {Ashoori}},\ }\bibfield  {title} {\bibinfo {title}
  {Massive {Dirac} {Fermions} and {Hofstadter} {Butterfly} in a van der {Waals}
  {Heterostructure}},\ }\href {https://doi.org/10.1126/science.1237240}
  {\bibfield  {journal} {\bibinfo  {journal} {Science}\ }\textbf {\bibinfo
  {volume} {340}},\ \bibinfo {pages} {1427} (\bibinfo {year}
  {2013})}\BibitemShut {NoStop}%
\bibitem [{\citenamefont {Dean}\ \emph {et~al.}(2013)\citenamefont {Dean},
  \citenamefont {Wang}, \citenamefont {Maher}, \citenamefont {Forsythe},
  \citenamefont {Ghahari}, \citenamefont {Gao}, \citenamefont {Katoch},
  \citenamefont {Ishigami}, \citenamefont {Moon}, \citenamefont {Koshino},
  \citenamefont {Taniguchi}, \citenamefont {Watanabe}, \citenamefont {Shepard},
  \citenamefont {Hone},\ and\ \citenamefont {Kim}}]{dean_hofstadter/s_2013}%
  \BibitemOpen
  \bibfield  {author} {\bibinfo {author} {\bibfnamefont {C.~R.}\ \bibnamefont
  {Dean}}, \bibinfo {author} {\bibfnamefont {L.}~\bibnamefont {Wang}}, \bibinfo
  {author} {\bibfnamefont {P.}~\bibnamefont {Maher}}, \bibinfo {author}
  {\bibfnamefont {C.}~\bibnamefont {Forsythe}}, \bibinfo {author}
  {\bibfnamefont {F.}~\bibnamefont {Ghahari}}, \bibinfo {author} {\bibfnamefont
  {Y.}~\bibnamefont {Gao}}, \bibinfo {author} {\bibfnamefont {J.}~\bibnamefont
  {Katoch}}, \bibinfo {author} {\bibfnamefont {M.}~\bibnamefont {Ishigami}},
  \bibinfo {author} {\bibfnamefont {P.}~\bibnamefont {Moon}}, \bibinfo {author}
  {\bibfnamefont {M.}~\bibnamefont {Koshino}}, \bibinfo {author} {\bibfnamefont
  {T.}~\bibnamefont {Taniguchi}}, \bibinfo {author} {\bibfnamefont
  {K.}~\bibnamefont {Watanabe}}, \bibinfo {author} {\bibfnamefont {K.~L.}\
  \bibnamefont {Shepard}}, \bibinfo {author} {\bibfnamefont {J.}~\bibnamefont
  {Hone}},\ and\ \bibinfo {author} {\bibfnamefont {P.}~\bibnamefont {Kim}},\
  }\bibfield  {title} {\bibinfo {title} {Hofstadter/'s butterfly and the
  fractal quantum {Hall} effect in moire superlattices},\ }\href
  {https://doi.org/10.1038/nature12186} {\bibfield  {journal} {\bibinfo
  {journal} {Nature}\ }\textbf {\bibinfo {volume} {497}},\ \bibinfo {pages}
  {598} (\bibinfo {year} {2013})}\BibitemShut {NoStop}%
\bibitem [{\citenamefont {Wallbank}\ \emph {et~al.}(2018)\citenamefont
  {Wallbank}, \citenamefont {Kumar}, \citenamefont {Holwill}, \citenamefont
  {Wang}, \citenamefont {Auton}, \citenamefont {Birkbeck}, \citenamefont
  {Mishchenko}, \citenamefont {Ponomarenko}, \citenamefont {Watanabe},
  \citenamefont {Taniguchi}, \citenamefont {Novoselov}, \citenamefont
  {Aleiner}, \citenamefont {Geim},\ and\ \citenamefont
  {Fal’ko}}]{wallbank_excess_2018}%
  \BibitemOpen
  \bibfield  {author} {\bibinfo {author} {\bibfnamefont {J.~R.}\ \bibnamefont
  {Wallbank}}, \bibinfo {author} {\bibfnamefont {R.~K.}\ \bibnamefont {Kumar}},
  \bibinfo {author} {\bibfnamefont {M.}~\bibnamefont {Holwill}}, \bibinfo
  {author} {\bibfnamefont {Z.}~\bibnamefont {Wang}}, \bibinfo {author}
  {\bibfnamefont {G.~H.}\ \bibnamefont {Auton}}, \bibinfo {author}
  {\bibfnamefont {J.}~\bibnamefont {Birkbeck}}, \bibinfo {author}
  {\bibfnamefont {A.}~\bibnamefont {Mishchenko}}, \bibinfo {author}
  {\bibfnamefont {L.~A.}\ \bibnamefont {Ponomarenko}}, \bibinfo {author}
  {\bibfnamefont {K.}~\bibnamefont {Watanabe}}, \bibinfo {author}
  {\bibfnamefont {T.}~\bibnamefont {Taniguchi}}, \bibinfo {author}
  {\bibfnamefont {K.~S.}\ \bibnamefont {Novoselov}}, \bibinfo {author}
  {\bibfnamefont {I.~L.}\ \bibnamefont {Aleiner}}, \bibinfo {author}
  {\bibfnamefont {A.~K.}\ \bibnamefont {Geim}},\ and\ \bibinfo {author}
  {\bibfnamefont {V.~I.}\ \bibnamefont {Fal’ko}},\ }\bibfield  {title}
  {\bibinfo {title} {Excess resistivity in graphene superlattices caused by
  umklapp electron–electron scattering},\ }\href
  {https://doi.org/10.1038/s41567-018-0278-6} {\bibfield  {journal} {\bibinfo
  {journal} {Nature Physics}\ ,\ \bibinfo {pages} {1}} (\bibinfo {year}
  {2018})}\BibitemShut {NoStop}%
\bibitem [{\citenamefont {Wang}\ \emph
  {et~al.}(2019{\natexlab{a}})\citenamefont {Wang}, \citenamefont {Wang},
  \citenamefont {Yin}, \citenamefont {Tovari}, \citenamefont {Yang},
  \citenamefont {Lin}, \citenamefont {Holwill}, \citenamefont {Birkbeck},
  \citenamefont {Perello}, \citenamefont {Xu}, \citenamefont {Zultak},
  \citenamefont {Gorbachev}, \citenamefont {Kretinin}, \citenamefont
  {Taniguchi}, \citenamefont {Watanabe}, \citenamefont {Morozov}, \citenamefont
  {Andelkovic}, \citenamefont {Milovanovic}, \citenamefont {Covaci},
  \citenamefont {Peeters}, \citenamefont {Mishchenko}, \citenamefont {Geim},
  \citenamefont {Novoselov}, \citenamefont {Fal'ko}, \citenamefont {Knothe},\
  and\ \citenamefont {Woods}}]{wang_composite_2019}%
  \BibitemOpen
  \bibfield  {author} {\bibinfo {author} {\bibfnamefont {Z.}~\bibnamefont
  {Wang}}, \bibinfo {author} {\bibfnamefont {Y.~B.}\ \bibnamefont {Wang}},
  \bibinfo {author} {\bibfnamefont {J.}~\bibnamefont {Yin}}, \bibinfo {author}
  {\bibfnamefont {E.}~\bibnamefont {Tovari}}, \bibinfo {author} {\bibfnamefont
  {Y.}~\bibnamefont {Yang}}, \bibinfo {author} {\bibfnamefont {L.}~\bibnamefont
  {Lin}}, \bibinfo {author} {\bibfnamefont {M.}~\bibnamefont {Holwill}},
  \bibinfo {author} {\bibfnamefont {J.}~\bibnamefont {Birkbeck}}, \bibinfo
  {author} {\bibfnamefont {D.~J.}\ \bibnamefont {Perello}}, \bibinfo {author}
  {\bibfnamefont {S.}~\bibnamefont {Xu}}, \bibinfo {author} {\bibfnamefont
  {J.}~\bibnamefont {Zultak}}, \bibinfo {author} {\bibfnamefont {R.~V.}\
  \bibnamefont {Gorbachev}}, \bibinfo {author} {\bibfnamefont {A.~V.}\
  \bibnamefont {Kretinin}}, \bibinfo {author} {\bibfnamefont {T.}~\bibnamefont
  {Taniguchi}}, \bibinfo {author} {\bibfnamefont {K.}~\bibnamefont {Watanabe}},
  \bibinfo {author} {\bibfnamefont {S.~V.}\ \bibnamefont {Morozov}}, \bibinfo
  {author} {\bibfnamefont {M.}~\bibnamefont {Andelkovic}}, \bibinfo {author}
  {\bibfnamefont {S.~P.}\ \bibnamefont {Milovanovic}}, \bibinfo {author}
  {\bibfnamefont {L.}~\bibnamefont {Covaci}}, \bibinfo {author} {\bibfnamefont
  {F.~M.}\ \bibnamefont {Peeters}}, \bibinfo {author} {\bibfnamefont
  {A.}~\bibnamefont {Mishchenko}}, \bibinfo {author} {\bibfnamefont {A.~K.}\
  \bibnamefont {Geim}}, \bibinfo {author} {\bibfnamefont {K.~S.}\ \bibnamefont
  {Novoselov}}, \bibinfo {author} {\bibfnamefont {V.~I.}\ \bibnamefont
  {Fal'ko}}, \bibinfo {author} {\bibfnamefont {A.}~\bibnamefont {Knothe}},\
  and\ \bibinfo {author} {\bibfnamefont {C.~R.}\ \bibnamefont {Woods}},\
  }\bibfield  {title} {\bibinfo {title} {Composite super-moire lattices in
  double-aligned graphene heterostructures},\ }\bibfield  {journal} {\bibinfo
  {journal} {Science Advances}\ }\textbf {\bibinfo {volume} {5}},\ \href
  {https://doi.org/10.1126/sciadv.aay8897} {10.1126/sciadv.aay8897} (\bibinfo
  {year} {2019}{\natexlab{a}})\BibitemShut {NoStop}%
\bibitem [{\citenamefont {Small}\ \emph {et~al.}(2003)\citenamefont {Small},
  \citenamefont {Shi},\ and\ \citenamefont {Kim}}]{small_mesoscopic_2003}%
  \BibitemOpen
  \bibfield  {author} {\bibinfo {author} {\bibfnamefont {J.~P.}\ \bibnamefont
  {Small}}, \bibinfo {author} {\bibfnamefont {L.}~\bibnamefont {Shi}},\ and\
  \bibinfo {author} {\bibfnamefont {P.}~\bibnamefont {Kim}},\ }\bibfield
  {title} {\bibinfo {title} {Mesoscopic thermal and thermoelectric measurements
  of individual carbon nanotubes},\ }\href
  {https://doi.org/10.1016/S0038-1098(03)00341-7} {\bibfield  {journal}
  {\bibinfo  {journal} {Solid State Commun.}\ }\bibinfo {series} {Quantum
  {Phases} at the {Nanoscale}},\ \textbf {\bibinfo {volume} {127}},\ \bibinfo
  {pages} {181} (\bibinfo {year} {2003})}\BibitemShut {NoStop}%
\bibitem [{\citenamefont {Guarochico-Moreira}\ \emph
  {et~al.}(2022)\citenamefont {Guarochico-Moreira}, \citenamefont {Sambricio},
  \citenamefont {Omari}, \citenamefont {Anderson}, \citenamefont {Bandurin},
  \citenamefont {Toscano-Figueroa}, \citenamefont {Natera-Cordero},
  \citenamefont {Watanabe}, \citenamefont {Taniguchi}, \citenamefont
  {Grigorieva},\ and\ \citenamefont {Vera-Marun}}]{Guarochico-Moreira2022}%
  \BibitemOpen
  \bibfield  {author} {\bibinfo {author} {\bibfnamefont {V.~H.}\ \bibnamefont
  {Guarochico-Moreira}}, \bibinfo {author} {\bibfnamefont {J.~L.}\ \bibnamefont
  {Sambricio}}, \bibinfo {author} {\bibfnamefont {K.}~\bibnamefont {Omari}},
  \bibinfo {author} {\bibfnamefont {C.~R.}\ \bibnamefont {Anderson}}, \bibinfo
  {author} {\bibfnamefont {D.~A.}\ \bibnamefont {Bandurin}}, \bibinfo {author}
  {\bibfnamefont {J.~C.}\ \bibnamefont {Toscano-Figueroa}}, \bibinfo {author}
  {\bibfnamefont {N.}~\bibnamefont {Natera-Cordero}}, \bibinfo {author}
  {\bibfnamefont {K.}~\bibnamefont {Watanabe}}, \bibinfo {author}
  {\bibfnamefont {T.}~\bibnamefont {Taniguchi}}, \bibinfo {author}
  {\bibfnamefont {I.~V.}\ \bibnamefont {Grigorieva}},\ and\ \bibinfo {author}
  {\bibfnamefont {I.~J.}\ \bibnamefont {Vera-Marun}},\ }\bibfield  {title}
  {\bibinfo {title} {{Tunable Spin Injection in High-Quality Graphene with
  One-Dimensional Contacts}},\ }\href
  {https://doi.org/10.1021/acs.nanolett.1c03625} {\bibfield  {journal}
  {\bibinfo  {journal} {Nano Letters}\ }\textbf {\bibinfo {volume} {22}},\
  \bibinfo {pages} {935} (\bibinfo {year} {2022})}\BibitemShut {NoStop}%
\bibitem [{\citenamefont {Ghahari}\ \emph
  {et~al.}(2016{\natexlab{b}})\citenamefont {Ghahari}, \citenamefont {Xie},
  \citenamefont {Taniguchi}, \citenamefont {Watanabe}, \citenamefont {Foster},\
  and\ \citenamefont {Kim}}]{ghahari_enhanced_2016}%
  \BibitemOpen
  \bibfield  {author} {\bibinfo {author} {\bibfnamefont {F.}~\bibnamefont
  {Ghahari}}, \bibinfo {author} {\bibfnamefont {H.-Y.}\ \bibnamefont {Xie}},
  \bibinfo {author} {\bibfnamefont {T.}~\bibnamefont {Taniguchi}}, \bibinfo
  {author} {\bibfnamefont {K.}~\bibnamefont {Watanabe}}, \bibinfo {author}
  {\bibfnamefont {M.~S.}\ \bibnamefont {Foster}},\ and\ \bibinfo {author}
  {\bibfnamefont {P.}~\bibnamefont {Kim}},\ }\bibfield  {title} {\bibinfo
  {title} {Enhanced {Thermoelectric} {Power} in {Graphene}: {Violation} of the
  {Mott} {Relation} by {Inelastic} {Scattering}},\ }\href
  {https://doi.org/10.1103/PhysRevLett.116.136802} {\bibfield  {journal}
  {\bibinfo  {journal} {Phys. Rev. Lett.}\ }\textbf {\bibinfo {volume} {116}},\
  \bibinfo {pages} {136802} (\bibinfo {year} {2016}{\natexlab{b}})}\BibitemShut
  {NoStop}%
\bibitem [{\citenamefont {Yankowitz}\ \emph {et~al.}(2012)\citenamefont
  {Yankowitz}, \citenamefont {Xue}, \citenamefont {Cormode}, \citenamefont
  {Sanchez-Yamagishi}, \citenamefont {Watanabe}, \citenamefont {Taniguchi},
  \citenamefont {Jarillo-Herrero}, \citenamefont {Jacquod},\ and\ \citenamefont
  {LeRoy}}]{yankowitz_emergence_2012}%
  \BibitemOpen
  \bibfield  {author} {\bibinfo {author} {\bibfnamefont {M.}~\bibnamefont
  {Yankowitz}}, \bibinfo {author} {\bibfnamefont {J.}~\bibnamefont {Xue}},
  \bibinfo {author} {\bibfnamefont {D.}~\bibnamefont {Cormode}}, \bibinfo
  {author} {\bibfnamefont {J.~D.}\ \bibnamefont {Sanchez-Yamagishi}}, \bibinfo
  {author} {\bibfnamefont {K.}~\bibnamefont {Watanabe}}, \bibinfo {author}
  {\bibfnamefont {T.}~\bibnamefont {Taniguchi}}, \bibinfo {author}
  {\bibfnamefont {P.}~\bibnamefont {Jarillo-Herrero}}, \bibinfo {author}
  {\bibfnamefont {P.}~\bibnamefont {Jacquod}},\ and\ \bibinfo {author}
  {\bibfnamefont {B.~J.}\ \bibnamefont {LeRoy}},\ }\bibfield  {title} {\bibinfo
  {title} {Emergence of superlattice {Dirac} points in graphene on hexagonal
  boron nitride},\ }\href {https://doi.org/10.1038/nphys2272} {\bibfield
  {journal} {\bibinfo  {journal} {Nature Physics}\ }\textbf {\bibinfo {volume}
  {8}},\ \bibinfo {pages} {382} (\bibinfo {year} {2012})}\BibitemShut {NoStop}%
\bibitem [{\citenamefont {Kim}\ \emph {et~al.}(2020)\citenamefont {Kim},
  \citenamefont {Xu}, \citenamefont {Berdyugin}, \citenamefont {Principi},
  \citenamefont {Slizovskiy}, \citenamefont {Xin}, \citenamefont
  {Kumaravadivel}, \citenamefont {Kuang}, \citenamefont {Hamer}, \citenamefont
  {Krishna~Kumar}, \citenamefont {Gorbachev}, \citenamefont {Watanabe},
  \citenamefont {Taniguchi}, \citenamefont {Grigorieva}, \citenamefont
  {Fal'ko}, \citenamefont {Polini},\ and\ \citenamefont
  {Geim}}]{proximity_screening_Geim}%
  \BibitemOpen
  \bibfield  {author} {\bibinfo {author} {\bibfnamefont {M.}~\bibnamefont
  {Kim}}, \bibinfo {author} {\bibfnamefont {S.~G.}\ \bibnamefont {Xu}},
  \bibinfo {author} {\bibfnamefont {A.~I.}\ \bibnamefont {Berdyugin}}, \bibinfo
  {author} {\bibfnamefont {A.}~\bibnamefont {Principi}}, \bibinfo {author}
  {\bibfnamefont {S.}~\bibnamefont {Slizovskiy}}, \bibinfo {author}
  {\bibfnamefont {N.}~\bibnamefont {Xin}}, \bibinfo {author} {\bibfnamefont
  {P.}~\bibnamefont {Kumaravadivel}}, \bibinfo {author} {\bibfnamefont
  {W.}~\bibnamefont {Kuang}}, \bibinfo {author} {\bibfnamefont
  {M.}~\bibnamefont {Hamer}}, \bibinfo {author} {\bibfnamefont
  {R.}~\bibnamefont {Krishna~Kumar}}, \bibinfo {author} {\bibfnamefont {R.~V.}\
  \bibnamefont {Gorbachev}}, \bibinfo {author} {\bibfnamefont {K.}~\bibnamefont
  {Watanabe}}, \bibinfo {author} {\bibfnamefont {T.}~\bibnamefont {Taniguchi}},
  \bibinfo {author} {\bibfnamefont {I.~V.}\ \bibnamefont {Grigorieva}},
  \bibinfo {author} {\bibfnamefont {V.~I.}\ \bibnamefont {Fal'ko}}, \bibinfo
  {author} {\bibfnamefont {M.}~\bibnamefont {Polini}},\ and\ \bibinfo {author}
  {\bibfnamefont {A.~K.}\ \bibnamefont {Geim}},\ }\bibfield  {title} {\bibinfo
  {title} {Control of electron-electron interaction in graphene by proximity
  screening},\ }\bibfield  {journal} {\bibinfo  {journal} {Nature
  Communications}\ }\textbf {\bibinfo {volume} {11}},\ \href
  {https://doi.org/10.1038/s41467-020-15829-1} {10.1038/s41467-020-15829-1}
  (\bibinfo {year} {2020})\BibitemShut {NoStop}%
\bibitem [{\citenamefont {Song}\ \emph {et~al.}(2013)\citenamefont {Song},
  \citenamefont {Shytov},\ and\ \citenamefont {Levitov}}]{song_electron_2013}%
  \BibitemOpen
  \bibfield  {author} {\bibinfo {author} {\bibfnamefont {J.~C.~W.}\
  \bibnamefont {Song}}, \bibinfo {author} {\bibfnamefont {A.~V.}\ \bibnamefont
  {Shytov}},\ and\ \bibinfo {author} {\bibfnamefont {L.~S.}\ \bibnamefont
  {Levitov}},\ }\bibfield  {title} {\bibinfo {title} {Electron {Interactions}
  and {Gap} {Opening} in {Graphene} {Superlattices}},\ }\href
  {https://doi.org/10.1103/PhysRevLett.111.266801} {\bibfield  {journal}
  {\bibinfo  {journal} {Phys. Rev. Lett.}\ }\textbf {\bibinfo {volume} {111}},\
  \bibinfo {pages} {266801} (\bibinfo {year} {2013})}\BibitemShut {NoStop}%
\bibitem [{\citenamefont {San-Jose}\ \emph
  {et~al.}(2014{\natexlab{a}})\citenamefont {San-Jose}, \citenamefont
  {Gutiérrez-Rubio}, \citenamefont {Sturla},\ and\ \citenamefont
  {Guinea}}]{san-jose_spontaneous_2014}%
  \BibitemOpen
  \bibfield  {author} {\bibinfo {author} {\bibfnamefont {P.}~\bibnamefont
  {San-Jose}}, \bibinfo {author} {\bibfnamefont {A.}~\bibnamefont
  {Gutiérrez-Rubio}}, \bibinfo {author} {\bibfnamefont {M.}~\bibnamefont
  {Sturla}},\ and\ \bibinfo {author} {\bibfnamefont {F.}~\bibnamefont
  {Guinea}},\ }\bibfield  {title} {\bibinfo {title} {Spontaneous strains and
  gap in graphene on boron nitride},\ }\href
  {https://doi.org/10.1103/PhysRevB.90.075428} {\bibfield  {journal} {\bibinfo
  {journal} {Phys. Rev. B}\ }\textbf {\bibinfo {volume} {90}},\ \bibinfo
  {pages} {075428} (\bibinfo {year} {2014}{\natexlab{a}})}\BibitemShut
  {NoStop}%
\bibitem [{\citenamefont {San-Jose}\ \emph
  {et~al.}(2014{\natexlab{b}})\citenamefont {San-Jose}, \citenamefont
  {Gutiérrez-Rubio}, \citenamefont {Sturla},\ and\ \citenamefont
  {Guinea}}]{san-jose_electronic_2014}%
  \BibitemOpen
  \bibfield  {author} {\bibinfo {author} {\bibfnamefont {P.}~\bibnamefont
  {San-Jose}}, \bibinfo {author} {\bibfnamefont {A.}~\bibnamefont
  {Gutiérrez-Rubio}}, \bibinfo {author} {\bibfnamefont {M.}~\bibnamefont
  {Sturla}},\ and\ \bibinfo {author} {\bibfnamefont {F.}~\bibnamefont
  {Guinea}},\ }\bibfield  {title} {\bibinfo {title} {Electronic structure of
  spontaneously strained graphene on hexagonal boron nitride},\ }\href
  {https://doi.org/10.1103/PhysRevB.90.115152} {\bibfield  {journal} {\bibinfo
  {journal} {Phys. Rev. B}\ }\textbf {\bibinfo {volume} {90}},\ \bibinfo
  {pages} {115152} (\bibinfo {year} {2014}{\natexlab{b}})}\BibitemShut
  {NoStop}%
\bibitem [{\citenamefont {Bokdam}\ \emph {et~al.}(2014)\citenamefont {Bokdam},
  \citenamefont {Amlaki}, \citenamefont {Brocks},\ and\ \citenamefont
  {Kelly}}]{bokdam_band_2014}%
  \BibitemOpen
  \bibfield  {author} {\bibinfo {author} {\bibfnamefont {M.}~\bibnamefont
  {Bokdam}}, \bibinfo {author} {\bibfnamefont {T.}~\bibnamefont {Amlaki}},
  \bibinfo {author} {\bibfnamefont {G.}~\bibnamefont {Brocks}},\ and\ \bibinfo
  {author} {\bibfnamefont {P.~J.}\ \bibnamefont {Kelly}},\ }\bibfield  {title}
  {\bibinfo {title} {Band gaps in incommensurable graphene on hexagonal boron
  nitride},\ }\href {https://doi.org/10.1103/PhysRevB.89.201404} {\bibfield
  {journal} {\bibinfo  {journal} {Phys. Rev. B}\ }\textbf {\bibinfo {volume}
  {89}},\ \bibinfo {pages} {201404} (\bibinfo {year} {2014})}\BibitemShut
  {NoStop}%
\bibitem [{\citenamefont {Jung}\ \emph {et~al.}(2015)\citenamefont {Jung},
  \citenamefont {DaSilva}, \citenamefont {MacDonald},\ and\ \citenamefont
  {Adam}}]{jung_origin_2015}%
  \BibitemOpen
  \bibfield  {author} {\bibinfo {author} {\bibfnamefont {J.}~\bibnamefont
  {Jung}}, \bibinfo {author} {\bibfnamefont {A.~M.}\ \bibnamefont {DaSilva}},
  \bibinfo {author} {\bibfnamefont {A.~H.}\ \bibnamefont {MacDonald}},\ and\
  \bibinfo {author} {\bibfnamefont {S.}~\bibnamefont {Adam}},\ }\bibfield
  {title} {\bibinfo {title} {Origin of band gaps in graphene on hexagonal boron
  nitride},\ }\bibfield  {journal} {\bibinfo  {journal} {Nat Commun}\ }\textbf
  {\bibinfo {volume} {6}},\ \href {https://doi.org/10.1038/ncomms7308}
  {10.1038/ncomms7308} (\bibinfo {year} {2015})\BibitemShut {NoStop}%
\bibitem [{\citenamefont {Moon}\ and\ \citenamefont
  {Koshino}(2014)}]{moon_electronic_2014}%
  \BibitemOpen
  \bibfield  {author} {\bibinfo {author} {\bibfnamefont {P.}~\bibnamefont
  {Moon}}\ and\ \bibinfo {author} {\bibfnamefont {M.}~\bibnamefont {Koshino}},\
  }\bibfield  {title} {\bibinfo {title} {Electronic properties of
  graphene/hexagonal-boron-nitride {M}oiré superlattice},\ }\href
  {https://doi.org/10.1103/PhysRevB.90.155406} {\bibfield  {journal} {\bibinfo
  {journal} {Phys. Rev. B}\ }\textbf {\bibinfo {volume} {90}},\ \bibinfo
  {pages} {155406} (\bibinfo {year} {2014})}\BibitemShut {NoStop}%
\bibitem [{\citenamefont {Gorbachev}\ \emph {et~al.}(2014)\citenamefont
  {Gorbachev}, \citenamefont {Song}, \citenamefont {Yu}, \citenamefont
  {Kretinin}, \citenamefont {Withers}, \citenamefont {Cao}, \citenamefont
  {Mishchenko}, \citenamefont {Grigorieva}, \citenamefont {Novoselov},
  \citenamefont {Levitov},\ and\ \citenamefont
  {Geim}}]{gorbachev_detecting_2014}%
  \BibitemOpen
  \bibfield  {author} {\bibinfo {author} {\bibfnamefont {R.~V.}\ \bibnamefont
  {Gorbachev}}, \bibinfo {author} {\bibfnamefont {J.~C.~W.}\ \bibnamefont
  {Song}}, \bibinfo {author} {\bibfnamefont {G.~L.}\ \bibnamefont {Yu}},
  \bibinfo {author} {\bibfnamefont {A.~V.}\ \bibnamefont {Kretinin}}, \bibinfo
  {author} {\bibfnamefont {F.}~\bibnamefont {Withers}}, \bibinfo {author}
  {\bibfnamefont {Y.}~\bibnamefont {Cao}}, \bibinfo {author} {\bibfnamefont
  {A.}~\bibnamefont {Mishchenko}}, \bibinfo {author} {\bibfnamefont {I.~V.}\
  \bibnamefont {Grigorieva}}, \bibinfo {author} {\bibfnamefont {K.~S.}\
  \bibnamefont {Novoselov}}, \bibinfo {author} {\bibfnamefont {L.~S.}\
  \bibnamefont {Levitov}},\ and\ \bibinfo {author} {\bibfnamefont {A.~K.}\
  \bibnamefont {Geim}},\ }\bibfield  {title} {\bibinfo {title} {Detecting
  topological currents in graphene superlattices},\ }\href
  {https://doi.org/10.1126/science.1254966} {\bibfield  {journal} {\bibinfo
  {journal} {Science}\ }\textbf {\bibinfo {volume} {346}},\ \bibinfo {pages}
  {448} (\bibinfo {year} {2014})}\BibitemShut {NoStop}%
\bibitem [{\citenamefont {Kretinin}\ \emph {et~al.}(2014)\citenamefont
  {Kretinin}, \citenamefont {Cao}, \citenamefont {Tu}, \citenamefont {Yu},
  \citenamefont {Jalil}, \citenamefont {Novoselov}, \citenamefont {Haigh},
  \citenamefont {Gholinia}, \citenamefont {Mishchenko}, \citenamefont {Lozada},
  \citenamefont {Georgiou}, \citenamefont {Woods}, \citenamefont {Withers},
  \citenamefont {Blake}, \citenamefont {Eda}, \citenamefont {Wirsig},
  \citenamefont {Hucho}, \citenamefont {Watanabe}, \citenamefont {Taniguchi},
  \citenamefont {Geim},\ and\ \citenamefont
  {Gorbachev}}]{kretinin_electronic_2014}%
  \BibitemOpen
  \bibfield  {author} {\bibinfo {author} {\bibfnamefont {A.~V.}\ \bibnamefont
  {Kretinin}}, \bibinfo {author} {\bibfnamefont {Y.}~\bibnamefont {Cao}},
  \bibinfo {author} {\bibfnamefont {J.~S.}\ \bibnamefont {Tu}}, \bibinfo
  {author} {\bibfnamefont {G.~L.}\ \bibnamefont {Yu}}, \bibinfo {author}
  {\bibfnamefont {R.}~\bibnamefont {Jalil}}, \bibinfo {author} {\bibfnamefont
  {K.~S.}\ \bibnamefont {Novoselov}}, \bibinfo {author} {\bibfnamefont {S.~J.}\
  \bibnamefont {Haigh}}, \bibinfo {author} {\bibfnamefont {A.}~\bibnamefont
  {Gholinia}}, \bibinfo {author} {\bibfnamefont {A.}~\bibnamefont
  {Mishchenko}}, \bibinfo {author} {\bibfnamefont {M.}~\bibnamefont {Lozada}},
  \bibinfo {author} {\bibfnamefont {T.}~\bibnamefont {Georgiou}}, \bibinfo
  {author} {\bibfnamefont {C.~R.}\ \bibnamefont {Woods}}, \bibinfo {author}
  {\bibfnamefont {F.}~\bibnamefont {Withers}}, \bibinfo {author} {\bibfnamefont
  {P.}~\bibnamefont {Blake}}, \bibinfo {author} {\bibfnamefont
  {G.}~\bibnamefont {Eda}}, \bibinfo {author} {\bibfnamefont {A.}~\bibnamefont
  {Wirsig}}, \bibinfo {author} {\bibfnamefont {C.}~\bibnamefont {Hucho}},
  \bibinfo {author} {\bibfnamefont {K.}~\bibnamefont {Watanabe}}, \bibinfo
  {author} {\bibfnamefont {T.}~\bibnamefont {Taniguchi}}, \bibinfo {author}
  {\bibfnamefont {A.~K.}\ \bibnamefont {Geim}},\ and\ \bibinfo {author}
  {\bibfnamefont {R.~V.}\ \bibnamefont {Gorbachev}},\ }\bibfield  {title}
  {\bibinfo {title} {Electronic {Properties} of {Graphene} {Encapsulated} with
  {Different} {Two}-{Dimensional} {Atomic} {Crystals}},\ }\href
  {https://doi.org/10.1021/nl5006542} {\bibfield  {journal} {\bibinfo
  {journal} {Nano Lett.}\ }\textbf {\bibinfo {volume} {14}},\ \bibinfo {pages}
  {3270} (\bibinfo {year} {2014})}\BibitemShut {NoStop}%
\bibitem [{\citenamefont {Martin}\ \emph {et~al.}(2008)\citenamefont {Martin},
  \citenamefont {Akerman}, \citenamefont {Ulbricht}, \citenamefont {Lohmann},
  \citenamefont {Smet}, \citenamefont {Von~Klitzing},\ and\ \citenamefont
  {Yacoby}}]{martin_observation_2008}%
  \BibitemOpen
  \bibfield  {author} {\bibinfo {author} {\bibfnamefont {J.}~\bibnamefont
  {Martin}}, \bibinfo {author} {\bibfnamefont {N.}~\bibnamefont {Akerman}},
  \bibinfo {author} {\bibfnamefont {G.}~\bibnamefont {Ulbricht}}, \bibinfo
  {author} {\bibfnamefont {T.}~\bibnamefont {Lohmann}}, \bibinfo {author}
  {\bibfnamefont {J.~H.}\ \bibnamefont {Smet}}, \bibinfo {author}
  {\bibfnamefont {K.}~\bibnamefont {Von~Klitzing}},\ and\ \bibinfo {author}
  {\bibfnamefont {A.}~\bibnamefont {Yacoby}},\ }\bibfield  {title} {\bibinfo
  {title} {Observation of electron-hole puddles in graphene using a scanning
  single-electron transistor},\ }\href {https://doi.org/10.1038/nphys781}
  {\bibfield  {journal} {\bibinfo  {journal} {Nature Physics}\ }\textbf
  {\bibinfo {volume} {4}},\ \bibinfo {pages} {144} (\bibinfo {year}
  {2008})}\BibitemShut {NoStop}%
\bibitem [{\citenamefont {Fang}\ \emph {et~al.}(2007)\citenamefont {Fang},
  \citenamefont {Konar}, \citenamefont {Xing},\ and\ \citenamefont
  {Jena}}]{fang_carrier_2007}%
  \BibitemOpen
  \bibfield  {author} {\bibinfo {author} {\bibfnamefont {T.}~\bibnamefont
  {Fang}}, \bibinfo {author} {\bibfnamefont {A.}~\bibnamefont {Konar}},
  \bibinfo {author} {\bibfnamefont {H.}~\bibnamefont {Xing}},\ and\ \bibinfo
  {author} {\bibfnamefont {D.}~\bibnamefont {Jena}},\ }\bibfield  {title}
  {\bibinfo {title} {Carrier statistics and quantum capacitance of graphene
  sheets and ribbons},\ }\href {https://doi.org/10.1063/1.2776887} {\bibfield
  {journal} {\bibinfo  {journal} {Applied Physics Letters}\ }\textbf {\bibinfo
  {volume} {91}},\ \bibinfo {pages} {092109} (\bibinfo {year}
  {2007})}\BibitemShut {NoStop}%
\bibitem [{\citenamefont {Yu}\ \emph {et~al.}(2014)\citenamefont {Yu},
  \citenamefont {Gorbachev}, \citenamefont {Tu}, \citenamefont {Kretinin},
  \citenamefont {Cao}, \citenamefont {Jalil}, \citenamefont {Withers},
  \citenamefont {Ponomarenko}, \citenamefont {Piot}, \citenamefont {Potemski},
  \citenamefont {Elias}, \citenamefont {Chen}, \citenamefont {Watanabe},
  \citenamefont {Taniguchi}, \citenamefont {Grigorieva}, \citenamefont
  {Novoselov}, \citenamefont {Fal'Ko}, \citenamefont {Geim},\ and\
  \citenamefont {Mishchenko}}]{Yu2014}%
  \BibitemOpen
  \bibfield  {author} {\bibinfo {author} {\bibfnamefont {G.~L.}\ \bibnamefont
  {Yu}}, \bibinfo {author} {\bibfnamefont {R.~V.}\ \bibnamefont {Gorbachev}},
  \bibinfo {author} {\bibfnamefont {J.~S.}\ \bibnamefont {Tu}}, \bibinfo
  {author} {\bibfnamefont {A.~V.}\ \bibnamefont {Kretinin}}, \bibinfo {author}
  {\bibfnamefont {Y.}~\bibnamefont {Cao}}, \bibinfo {author} {\bibfnamefont
  {R.}~\bibnamefont {Jalil}}, \bibinfo {author} {\bibfnamefont
  {F.}~\bibnamefont {Withers}}, \bibinfo {author} {\bibfnamefont {L.~A.}\
  \bibnamefont {Ponomarenko}}, \bibinfo {author} {\bibfnamefont {B.~A.}\
  \bibnamefont {Piot}}, \bibinfo {author} {\bibfnamefont {M.}~\bibnamefont
  {Potemski}}, \bibinfo {author} {\bibfnamefont {D.~C.}\ \bibnamefont {Elias}},
  \bibinfo {author} {\bibfnamefont {X.}~\bibnamefont {Chen}}, \bibinfo {author}
  {\bibfnamefont {K.}~\bibnamefont {Watanabe}}, \bibinfo {author}
  {\bibfnamefont {T.}~\bibnamefont {Taniguchi}}, \bibinfo {author}
  {\bibfnamefont {I.~V.}\ \bibnamefont {Grigorieva}}, \bibinfo {author}
  {\bibfnamefont {K.~S.}\ \bibnamefont {Novoselov}}, \bibinfo {author}
  {\bibfnamefont {V.~I.}\ \bibnamefont {Fal'Ko}}, \bibinfo {author}
  {\bibfnamefont {A.~K.}\ \bibnamefont {Geim}},\ and\ \bibinfo {author}
  {\bibfnamefont {A.}~\bibnamefont {Mishchenko}},\ }\bibfield  {title}
  {\bibinfo {title} {{Hierarchy of Hofstadter states and replica quantum Hall
  ferromagnetism in graphene superlattices}},\ }\href
  {https://doi.org/10.1038/nphys2979} {\bibfield  {journal} {\bibinfo
  {journal} {Nature Physics}\ }\textbf {\bibinfo {volume} {10}},\ \bibinfo
  {pages} {525} (\bibinfo {year} {2014})}\BibitemShut {NoStop}%
\bibitem [{\citenamefont {Hwang}\ \emph {et~al.}(2009)\citenamefont {Hwang},
  \citenamefont {Rossi},\ and\ \citenamefont {Das~Sarma}}]{hwang_theory_2009}%
  \BibitemOpen
  \bibfield  {author} {\bibinfo {author} {\bibfnamefont {E.~H.}\ \bibnamefont
  {Hwang}}, \bibinfo {author} {\bibfnamefont {E.}~\bibnamefont {Rossi}},\ and\
  \bibinfo {author} {\bibfnamefont {S.}~\bibnamefont {Das~Sarma}},\ }\bibfield
  {title} {\bibinfo {title} {Theory of thermopower in two-dimensional
  graphene},\ }\href {https://doi.org/10.1103/PhysRevB.80.235415} {\bibfield
  {journal} {\bibinfo  {journal} {Phys. Rev. B}\ }\textbf {\bibinfo {volume}
  {80}},\ \bibinfo {pages} {235415} (\bibinfo {year} {2009})}\BibitemShut
  {NoStop}%
\bibitem [{\citenamefont {Hwang}\ \emph {et~al.}(2007)\citenamefont {Hwang},
  \citenamefont {Adam},\ and\ \citenamefont {Sarma}}]{hwang_carrier_2007}%
  \BibitemOpen
  \bibfield  {author} {\bibinfo {author} {\bibfnamefont {E.~H.}\ \bibnamefont
  {Hwang}}, \bibinfo {author} {\bibfnamefont {S.}~\bibnamefont {Adam}},\ and\
  \bibinfo {author} {\bibfnamefont {S.~D.}\ \bibnamefont {Sarma}},\ }\bibfield
  {title} {\bibinfo {title} {Carrier transport in two-dimensional graphene
  layers},\ }\href {https://doi.org/10.1103/PhysRevLett.98.186806} {\bibfield
  {journal} {\bibinfo  {journal} {Phys. Rev. Lett.}\ }\textbf {\bibinfo
  {volume} {98}},\ \bibinfo {pages} {186806} (\bibinfo {year}
  {2007})}\BibitemShut {NoStop}%
\bibitem [{\citenamefont {Dean}\ \emph {et~al.}(2010)\citenamefont {Dean},
  \citenamefont {Young}, \citenamefont {Meric}, \citenamefont {Lee},
  \citenamefont {Wang}, \citenamefont {Sorgenfrei}, \citenamefont {Watanabe},
  \citenamefont {Taniguchi}, \citenamefont {Kim}, \citenamefont {Shepard},\
  and\ \citenamefont {Hone}}]{dean_boron_2010}%
  \BibitemOpen
  \bibfield  {author} {\bibinfo {author} {\bibfnamefont {C.~R.}\ \bibnamefont
  {Dean}}, \bibinfo {author} {\bibfnamefont {A.~F.}\ \bibnamefont {Young}},
  \bibinfo {author} {\bibfnamefont {I.}~\bibnamefont {Meric}}, \bibinfo
  {author} {\bibfnamefont {C.}~\bibnamefont {Lee}}, \bibinfo {author}
  {\bibfnamefont {L.}~\bibnamefont {Wang}}, \bibinfo {author} {\bibfnamefont
  {S.}~\bibnamefont {Sorgenfrei}}, \bibinfo {author} {\bibfnamefont
  {K.}~\bibnamefont {Watanabe}}, \bibinfo {author} {\bibfnamefont
  {T.}~\bibnamefont {Taniguchi}}, \bibinfo {author} {\bibfnamefont
  {P.}~\bibnamefont {Kim}}, \bibinfo {author} {\bibfnamefont {K.~L.}\
  \bibnamefont {Shepard}},\ and\ \bibinfo {author} {\bibfnamefont
  {J.}~\bibnamefont {Hone}},\ }\bibfield  {title} {\bibinfo {title} {Boron
  nitride substrates for high-quality graphene electronics},\ }\href
  {https://doi.org/10.1038/nnano.2010.172} {\bibfield  {journal} {\bibinfo
  {journal} {Nature Nanotechnology}\ }\textbf {\bibinfo {volume} {5}},\
  \bibinfo {pages} {722} (\bibinfo {year} {2010})}\BibitemShut {NoStop}%
\bibitem [{\citenamefont {Wang}\ \emph
  {et~al.}(2019{\natexlab{b}})\citenamefont {Wang}, \citenamefont {Zihlmann},
  \citenamefont {Liu}, \citenamefont {Makk}, \citenamefont {Watanabe},
  \citenamefont {Taniguchi}, \citenamefont {Baumgartner},\ and\ \citenamefont
  {Schönenberger}}]{wang_new_generation_2019}%
  \BibitemOpen
  \bibfield  {author} {\bibinfo {author} {\bibfnamefont {L.}~\bibnamefont
  {Wang}}, \bibinfo {author} {\bibfnamefont {S.}~\bibnamefont {Zihlmann}},
  \bibinfo {author} {\bibfnamefont {M.-H.}\ \bibnamefont {Liu}}, \bibinfo
  {author} {\bibfnamefont {P.}~\bibnamefont {Makk}}, \bibinfo {author}
  {\bibfnamefont {K.}~\bibnamefont {Watanabe}}, \bibinfo {author}
  {\bibfnamefont {T.}~\bibnamefont {Taniguchi}}, \bibinfo {author}
  {\bibfnamefont {A.}~\bibnamefont {Baumgartner}},\ and\ \bibinfo {author}
  {\bibfnamefont {C.}~\bibnamefont {Schönenberger}},\ }\bibfield  {title}
  {\bibinfo {title} {New generation of moiré superlattices in doubly aligned
  hbn/graphene/hbn heterostructures},\ }\href
  {https://doi.org/10.1021/acs.nanolett.8b05061} {\bibfield  {journal}
  {\bibinfo  {journal} {Nano Letters}\ }\textbf {\bibinfo {volume} {19}},\
  \bibinfo {pages} {2371} (\bibinfo {year} {2019}{\natexlab{b}})}\BibitemShut
  {NoStop}%
\bibitem [{\citenamefont {Sun}\ \emph {et~al.}(2021)\citenamefont {Sun},
  \citenamefont {Zhang}, \citenamefont {Liu}, \citenamefont {Zhu},
  \citenamefont {Huang}, \citenamefont {Yuan}, \citenamefont {Wang},
  \citenamefont {Watanabe}, \citenamefont {Taniguchi}, \citenamefont {Li},
  \citenamefont {Zhu}, \citenamefont {Mao}, \citenamefont {Yang}, \citenamefont
  {Kang}, \citenamefont {Liu}, \citenamefont {Ye}, \citenamefont {Han},\ and\
  \citenamefont {Zhang}}]{sun_correlated_2021}%
  \BibitemOpen
  \bibfield  {author} {\bibinfo {author} {\bibfnamefont {X.}~\bibnamefont
  {Sun}}, \bibinfo {author} {\bibfnamefont {S.}~\bibnamefont {Zhang}}, \bibinfo
  {author} {\bibfnamefont {Z.}~\bibnamefont {Liu}}, \bibinfo {author}
  {\bibfnamefont {H.}~\bibnamefont {Zhu}}, \bibinfo {author} {\bibfnamefont
  {J.}~\bibnamefont {Huang}}, \bibinfo {author} {\bibfnamefont
  {K.}~\bibnamefont {Yuan}}, \bibinfo {author} {\bibfnamefont {Z.}~\bibnamefont
  {Wang}}, \bibinfo {author} {\bibfnamefont {K.}~\bibnamefont {Watanabe}},
  \bibinfo {author} {\bibfnamefont {T.}~\bibnamefont {Taniguchi}}, \bibinfo
  {author} {\bibfnamefont {X.}~\bibnamefont {Li}}, \bibinfo {author}
  {\bibfnamefont {M.}~\bibnamefont {Zhu}}, \bibinfo {author} {\bibfnamefont
  {J.}~\bibnamefont {Mao}}, \bibinfo {author} {\bibfnamefont {T.}~\bibnamefont
  {Yang}}, \bibinfo {author} {\bibfnamefont {J.}~\bibnamefont {Kang}}, \bibinfo
  {author} {\bibfnamefont {J.}~\bibnamefont {Liu}}, \bibinfo {author}
  {\bibfnamefont {Y.}~\bibnamefont {Ye}}, \bibinfo {author} {\bibfnamefont
  {Z.~V.}\ \bibnamefont {Han}},\ and\ \bibinfo {author} {\bibfnamefont
  {Z.}~\bibnamefont {Zhang}},\ }\bibfield  {title} {\bibinfo {title}
  {Correlated states in doubly-aligned hbn/graphene/hbn heterostructures},\
  }\bibfield  {journal} {\bibinfo  {journal} {Nature Communications}\ }\textbf
  {\bibinfo {volume} {12}},\ \href {https://doi.org/10.1038/s41467-021-27514-y}
  {10.1038/s41467-021-27514-y} (\bibinfo {year} {2021})\BibitemShut {NoStop}%
\bibitem [{\citenamefont {Issi}\ and\ \citenamefont {Boxus}(1979)}]{Issi1979}%
  \BibitemOpen
  \bibfield  {author} {\bibinfo {author} {\bibfnamefont {J.}~\bibnamefont
  {Issi}}\ and\ \bibinfo {author} {\bibfnamefont {J.}~\bibnamefont {Boxus}},\
  }\bibfield  {title} {\bibinfo {title} {{Phonon-drag low temperature
  thermoelectric refrigeration}},\ }\href
  {https://doi.org/10.1016/0011-2275(79)90004-3} {\bibfield  {journal}
  {\bibinfo  {journal} {Cryogenics}\ }\textbf {\bibinfo {volume} {19}},\
  \bibinfo {pages} {517} (\bibinfo {year} {1979})}\BibitemShut {NoStop}%
\bibitem [{\citenamefont {Cusack}\ and\ \citenamefont
  {Kendall}(1958)}]{Cusack1958}%
  \BibitemOpen
  \bibfield  {author} {\bibinfo {author} {\bibfnamefont {N.}~\bibnamefont
  {Cusack}}\ and\ \bibinfo {author} {\bibfnamefont {P.}~\bibnamefont
  {Kendall}},\ }\bibfield  {title} {\bibinfo {title} {{The Absolute Scale of
  Thermoelectric Power at High Temperature}},\ }\href
  {https://doi.org/10.1088/0370-1328/72/5/429} {\bibfield  {journal} {\bibinfo
  {journal} {Proceedings of the Physical Society}\ }\textbf {\bibinfo {volume}
  {72}},\ \bibinfo {pages} {898} (\bibinfo {year} {1958})}\BibitemShut
  {NoStop}%
\bibitem [{\citenamefont {Nam}\ \emph {et~al.}(2010{\natexlab{b}})\citenamefont
  {Nam}, \citenamefont {Ki},\ and\ \citenamefont {Lee}}]{Nam2010}%
  \BibitemOpen
  \bibfield  {author} {\bibinfo {author} {\bibfnamefont {S.-G.}\ \bibnamefont
  {Nam}}, \bibinfo {author} {\bibfnamefont {D.-K.}\ \bibnamefont {Ki}},\ and\
  \bibinfo {author} {\bibfnamefont {H.-J.}\ \bibnamefont {Lee}},\ }\bibfield
  {title} {\bibinfo {title} {{Thermoelectric transport of massive Dirac
  fermions in bilayer graphene}},\ }\href
  {https://doi.org/10.1103/PhysRevB.82.245416} {\bibfield  {journal} {\bibinfo
  {journal} {Physical Review B}\ }\textbf {\bibinfo {volume} {82}},\ \bibinfo
  {pages} {245416} (\bibinfo {year} {2010}{\natexlab{b}})}\BibitemShut
  {NoStop}%
\end{thebibliography}

%

\end{document}